\documentclass[sn-nature,Numbered]{sn-jnl}
\usepackage{enumitem}
\usepackage{graphicx}%
\usepackage{multirow}%
\usepackage{amsmath,amssymb,amsfonts}%
\usepackage{amsthm}%
\usepackage{mathrsfs}%
\usepackage[title]{appendix}%
\usepackage{xcolor}%
\usepackage{textcomp}%
\usepackage{manyfoot}%
\usepackage{soul}

\usepackage{xcolor}
\newcommand{\kolr}[1]{{{\color{black}{#1}}}}
\newcommand{\kolb}[1]{{{\color{black}{#1}}}}
\newcommand{\kolg}[1]{{{\color{black}{#1}}}}

\def\mc{infinitely thin ring}
\def\mcs{infinitely thin rings}
\def\potmc{\varphi_\text{itr}}
\def\den{{\sigma}}

\begin{document}

\title[The outer potential of an inhomogeneous torus]{The outer gravitational potential of an inhomogeneous torus with an elliptical cross-section}

\author*[1,2,3]{\fnm{Elena} \sur{Bannikova}}\email{olena.bannikova@inaf.it}

\author[2,3]{\fnm{Serhii} \sur{Skolota}}\email{sergeyskolota@gmail.com}
\equalcont{These authors contributed equally to this work.}

\author[1,4]{\fnm{Massimo} \sur{Capaccioli}}\email{massimo.capaccioli@inaf.it}
\equalcont{These authors contributed equally to this work.}

\affil*[1]{\orgname{INAF - Astronomical Observatory of Capodimonte}, \orgaddress{\street{Salita Moiariello 16}, \city{Naples}, \postcode{I-80131},  \country{Italy}}}

\affil[2]{\orgdiv{Institute of Radio Astronomy}, \orgname{National Academy of Sciences of Ukraine}, \orgaddress{\street{Mystetstv 4}, \city{Kharkiv}, \postcode{UA-61002}, \country{Ukraine}}}

\affil[3]{\orgname{V.N.Karazin Kharkiv National University}, \orgaddress{\street{Svobody Sq.4}, \city{Kharkiv}, \postcode{UA-61022}, \state{{State}}, \country{Ukraine}}}

\affil[4]{\orgname{University of Naples {\it Federico II}}, \orgaddress{\city{Naples}, \postcode{I-80126}, \country{Italy}}}

\abstract{Toroidal/ring structures are a common feature in a wide variety of astrophysical objects, including dusty tori in active galactic nuclei (AGNs), rings in galaxies, protoplanetary disks, and others. The matter distribution in such structures is not homogeneous and can be flattened by self-gravity or become elongated in the vertical direction, as is the case with obscuring tori in AGNs. This led us to consider the more general case of the gravitational  potential of an inhomogeneous torus with an elliptical cross-section. 
We begin by showing that the outer potential of a homogeneous elliptical torus can be effectively approximated with less than 1\% error by 
the potentials of two \mcs \,  with a minor correction term. These two rings have masses each equal to half the total mass of the torus. The most notable feature is that each such infinitely thin ring is positioned at precisely the halfway point between the center and the focus of the elliptical cross-section, regardless of the torus' other parameters. The result, which holds for both oblate and prolate geometries, allows us to find a new expression to handle the outer potential of an inhomogeneous torus with an elliptical cross-section. 
The confocal density distribution is a special case. We have found that the outer potential of such a torus is only weakly  dependent of the density distribution law.  Consequently, even for the confocal inhomogeneous case, the outer potential is well represented by two \mcs. This approach simplifies problems of dynamics and allows for the analysis of the results of N-body simulations for the systems consisting of toroidal structures. For completeness, we have derived the expressions for the components of the external force exerted by a homogeneous torus with an elliptical cross-section, both for the exact form of the potential and for our approximation by two infinitely thin rings. Comparison of the two shows that our model fits the true trend of the force well.}


\maketitle

\section{Introduction}\label{section1}\label{Intro} 

The gravitational potential of toroidal/ring structures is an essential tool for understanding the dynamics of a vast array of astrophysical objects: ring galaxies \citep{2011MNRAS.418.1834F,2018MNRAS.476.3269B,Alberti2019}, protostar systems \citep{1992ApJ...394..248W,Tresaco2011}, dusty tori in active galactic nuclei (AGNs) \citep{1985ApJ...297..621A, 1993ARA&A..31..473A, 1995PASP..107..803U}, \kolr{ and others}. Within the Solar System, the torus/ring potential has been used to investigate the Kuiper-Belt gravitational forces \citep{2005PhRvD..72h3004N}  { and to model the main asteroid belt \citep{2022Icar..37614845L}.
It can} be also applied to simulate the gravitational field of a ring-like solid mass distribution shaping the  surfaces of asteroids, such as that of the asteroid Bennu \citep{2020SciA....6.3350S}.  
Dusty tori have been directly observed in AGNs at different wavelength bands. Their masses are about 1 percent of their central supermassive black holes \citep{Jaffe2004,  2018ApJ...853L..25I, 2019A&A...632A..61G, Combes2019, 2019ApJ...874L..32C}. They also have geometrically thick structures with cross-sections elongated along the vertical direction.  
N-body simulations of a clumpy torus in the field of the central mass show
that the torus is geometrically thick with an elliptical cross-section and with a Gaussian density distribution within it \citep{2012MNRAS.424..820B, 2021MNRAS.503.1459B}.
This led us to consider the more general case of the gravitational potential of an inhomogeneous torus with an elliptical cross-section for both oblate and prolate configurations (that is, when either the major or the minor axis of the elliptical cross-section is aligned with the equatorial plane of the torus).

The potential of a torus can be obtained by direct integration over the volume, but the resulting expression is hard to use.
A way out is to express it by the sum of the potentials of simpler structures, as was done for the torus with the circular cross-section \cite{2011MNRAS.411..557B}. 
Since the limiting case of such a torus is that of zero
cross-section
and the structure degenerates into an \mc,
it was an obvious choice to adopt the \mc \, as the elementary component for building the torus potential. 
By expanding the potential in a Maclaurin series up to the 2nd term (in the vicinity of the central \mc), we have proved \cite{2011MNRAS.411..557B} that { the outer potential\footnote{We define the outer region of the torus as the region external to the torus body where the potential satisfies the Laplace equation.}} 
of a homogeneous torus with a circular cross-section is well { matched} by that of an \mc \, with the same mass located at the cross-section center. The same method has been used to obtain the outer potential of a thin circular toroidal shell \cite{2020MNRAS.494.5825H}. 

A different approach is based on the use of toroidal coordinates. This method was first applied in \cite{1973AnPhy..77..279W}, where the expansion of the electrostatic potential of a homogeneous torus with circular cross-section was considered. This representation was later generalized to the gravitational potential \cite{Majic2020}.  Another approach has been the numerical simulation of the gravitational potential as the toroidal series for the case of oval and Brillouin cross-sections  \cite{2016AJ....152...35F}. 
Despite all these investigations, there are only special analytical approximations for the gravitational potential of a torus with a circular cross-section and a homogeneous density distribution. This fact motivated the present study.
 
Our main goal is to find a convenient expression of the outer torus potential for the cases of elliptical cross-sections (both oblate and prolate)
and inhomogeneous density distributions, to be used for analytical investigations of dynamical problems and for discovering new gravitational properties of such structures.
To this end, we first consider the case of the potential of a torus with an elliptical cross-section and a homogeneous density distribution (Section~\ref{section2}). By using a model based on two \mcs, 
we find an approximate expression for the outer potential 
(Section~\ref{section3}) for arbitrary axis ratios in both the oblate and prolate cases (flattened/elongated tori). In Section~\ref{section4} we use this approach to obtain a new expression of the potential for an inhomogeneous torus and also to show that, for a torus with confocal isodensity contours, the potential is (almost) independent of the density law.

\section{Homogeneous torus with elliptical cross-section}\label{section2}
\begin{figure}
\centering
\includegraphics[width = 90mm]{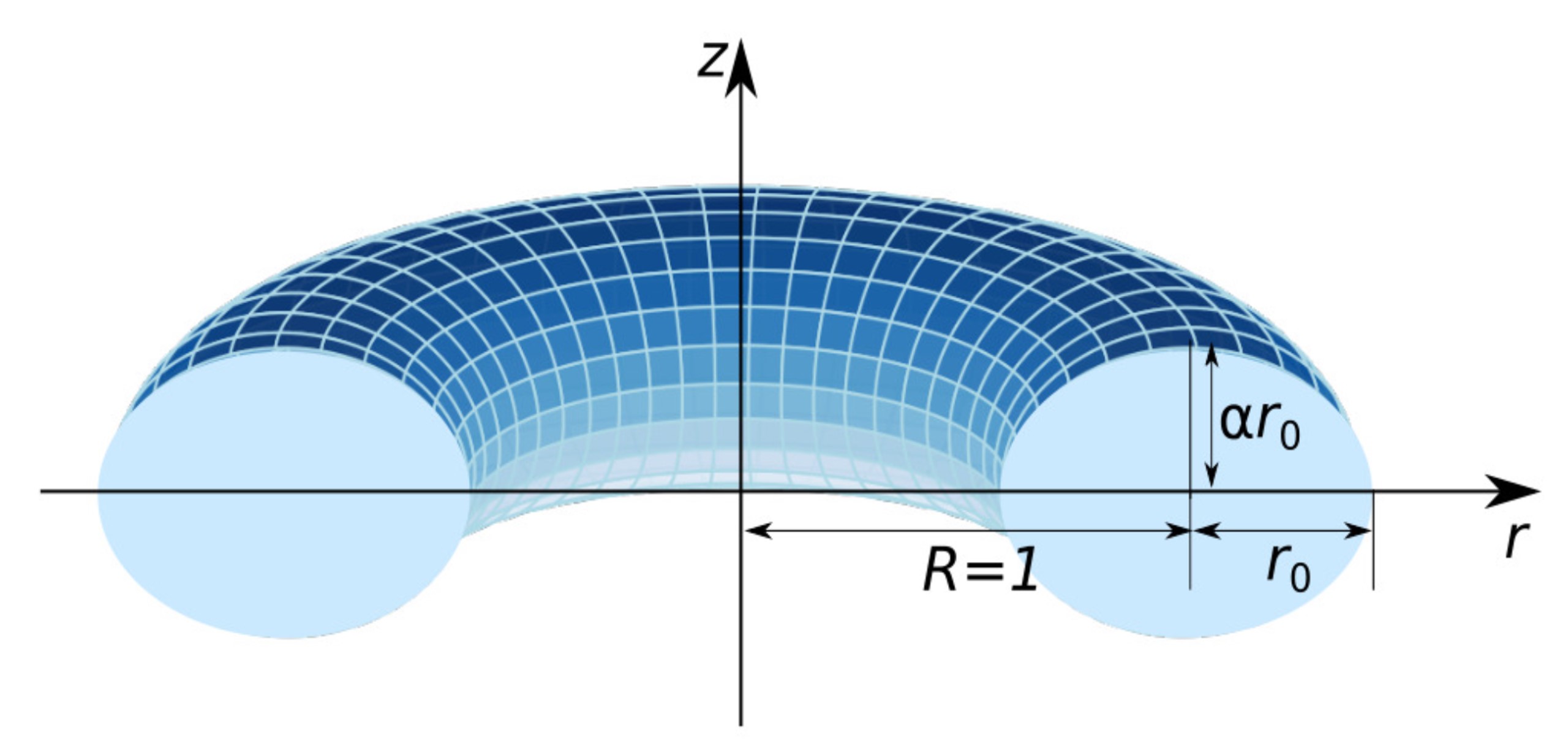}
\includegraphics[width = 65mm]{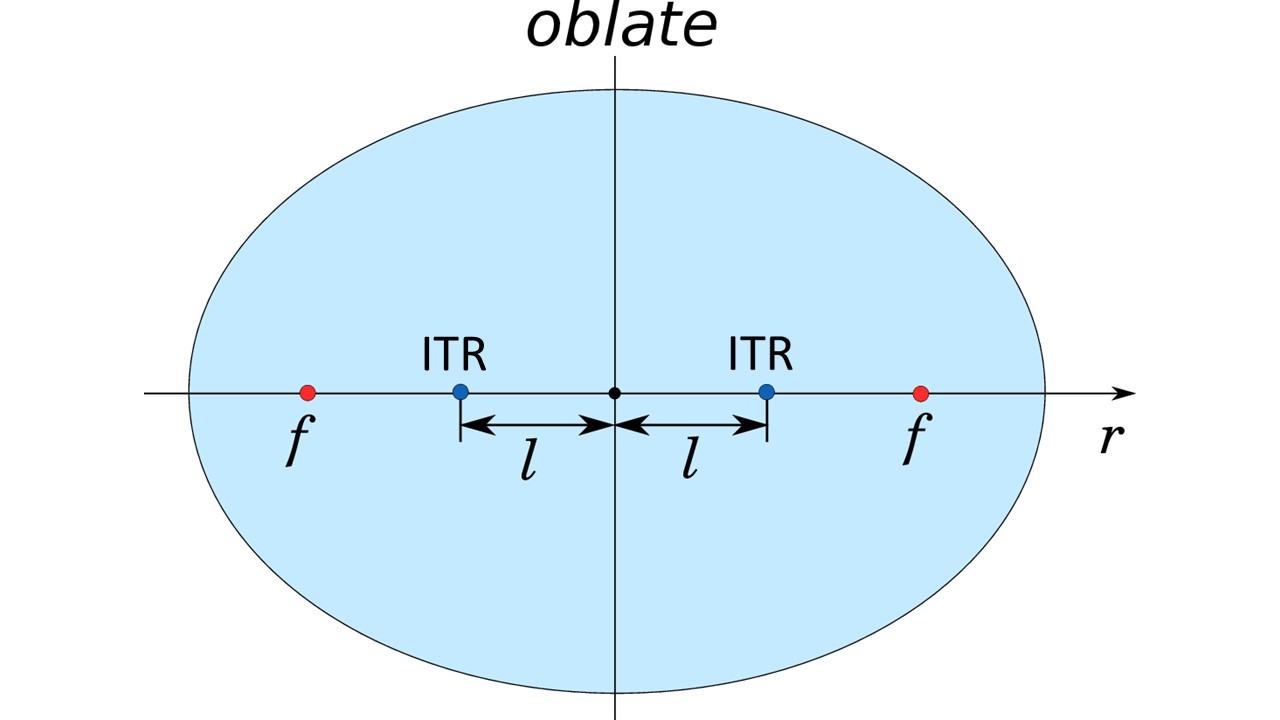}\qquad 
\includegraphics[width = 41mm]{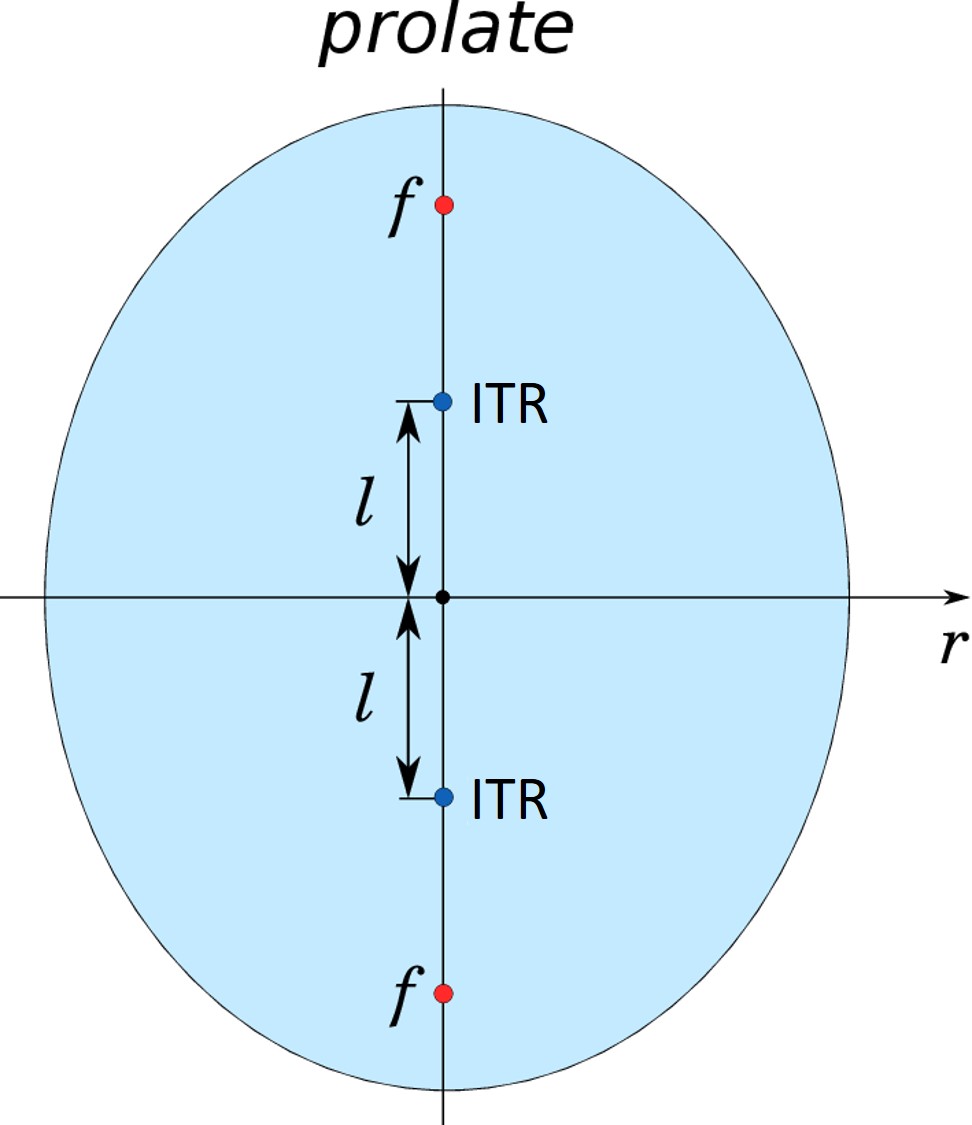}
\caption{{\it Top}: scheme of a solid torus with an elliptical cross-section (drawn as oblate in the present case). The cylindrical coordinates ($r,z$) are dimensionless and normalized to $R$. {\it Bottom}: scheme of tori with  oblate (left) and prolate (right) cross-sections. The blue dots mark the locations of the \mcs\, (labelled ITR for "Infinitely Thin Ring"\, in the figure) and the red dots are the foci.}
\label{fig:scheme}
\end{figure}

Consider a homogeneous torus with mass $M$ and major radius $R$ (Fig. \ref{fig:scheme}; {\it top}). We use dimensionless cylindrical coordinates which are normalized to $R$: \mbox{$r\rightarrow r/R$}, \mbox{$z\rightarrow z/R$}. The origin of the reference system coincides with the center of the torus and the $z$-axis is perpendicular to its equatorial plane. The torus cross-section is an ellipse with { dimensionless} semi-axes $a=r_0=R_0/R$ and $b = \alpha r_0=\alpha R_0/R$ (elliptical torus); 
$R_0$ is a dimensional term. The axis ratio $\alpha $ determines the degree of flattening/elongation of the torus
for both the oblate ($0<\alpha\le 1$; Fig.~\ref{fig:scheme}, bottom-left) and the prolate ($\alpha>1$; Fig.~\ref{fig:scheme}, bottom-right) cross-section.
The equation of the cross-section is
\begin{equation}\label{eq2.1}
\frac{(r -1)^2}{r_0^2}+\frac{z^2}{\alpha^2 r_0^2}=1.
\end{equation} 

\subsection{The exact expression of the potential}\label{formersection3}
We will derive the gravitational potential at a point $P(r, z)$ outside the torus body by generalizing the expression for the potential of a homogeneous torus with a circular cross-section \citep{2011MNRAS.411..557B} to the elliptical case using a set of massive \mcs. The final expression \citep{2021MNRAS.503.1459B} is:  
\begin{equation}\label{eq2.2}
\varphi_\text{tor} (r, z) = 
	\den_0\int\limits_{-r_0}^{r_0} 
        \int\limits_{-z_e}^{z_e}
             \potmc(r,z; x', z')dz'dx',
\end{equation}  
where $z_e = \alpha \sqrt{r_0^2 - x'^2}$.
Here the surface density of the torus cross-section per unit mass\footnote{Note that the dimensional factor $GM/(\pi R)$ is inside of $\potmc$ (see expression (\ref{eq2.3})).} is  \mbox{$\den_0 = 1/(\pi \alpha r_0^2) = \text{const}$}, while
 $\potmc$ is the potential of the 
 \mc \, with coordinates $(x',z')$ counted from the center of the torus cross-section\footnote{{ The semicolon here and in the following formulae separates the coordinates  from the parameter(s).}}:
\begin{equation}\label{eq2.3}
  \potmc (r,z;x',z') = \frac{G M}{\pi R}\frac{2(1 + x')}{\sqrt{(1+x'+r)^2 + (z - z')^2}}  \, K(m).
\end{equation} 
The function:
\begin{equation}\label{eq2.3b}
 K(m) = \int\limits_0^{\pi/2}d\beta / \sqrt{1-m\sin^2 \beta}   
\end{equation}
is the complete elliptical integral of the first kind with the parameter
\begin{equation}\label{eq2.4}
  \displaystyle m = \frac{4 r \, (1 + x')}{(1+ x'+ r)^2 +
  (z-z')^2}~,
\end{equation}
and $G$ is the gravitational constant.    
The expression (\ref{eq2.2}) holds at any point, both outside and inside the torus volume. It differs from the circular case only in that it includes the axis ratio $\alpha$, which appears in the limits $\pm z_e$ of the integral and in the density $\sigma_0$. As for the circular case, it cannot be integrated analytically. In the next Section we present an approximation for this potential which will prove useful for deriving some new properties to tackle the inhomogeneous density distributions.

\subsection{Approximation of the outer potential by two \mcs}\label{section3}
\subsubsection{General case}\label{section3a}
To find a simpler expression for the outer potential of a homogeneous torus with an elliptical cross-section, we use the solution obtained for a circular torus, when $\alpha =1$. 
Replacing the expression (\ref{eq2.3}) with its Maclaurin series in $x'$ and $z'$ up to second order, we obtain after integration
\citep{2011MNRAS.411..557B}\footnote{A similar expression was subsequently obtained in  \citep{2018JTePh..63..311K, 2020MNRAS.494.5825H} using the same method. For example, the formula (13), together with (14) in \citep{2018JTePh..63..311K} and (52), (54) in \citep{2020MNRAS.494.5825H}, becomes identical to formula (11) in \citep{2011MNRAS.411..557B} after some transformations and appropriate renaming. 
It should be noted that a misprint exists in the expression for the $S$-function in the original paper \citep{2011MNRAS.411..557B}. It can be simply detected repeating the analytical calculations. The correction of this misprint was made in the astro-ph version of the paper (arXiv:1009.4324) and cited in the book \citep{FCMbook}.}:
\begin{equation}\label{eq3.1}
\varphi_\text{tor} (r, z) \approx \varphi_{0}(r,z)\Big(1 -r_0^2 S(r,z)\Big).
\end{equation}   
Here the potential of the central \mc \, \kolb{\citep{Kellogg1929, 1983CeMec..30..225L}} is
\begin{equation}\label{eq3.2}
\varphi_{0}(r,z) = \frac{G M}{\pi R}\frac{2}{\sqrt{(1+r)^2 + z^2}}  \, K(m_0),
\end{equation} 
\kolr{ and}
\begin{equation}\label{eq3.3}
 S(r,z) = \frac{1}{16}\left(1 - \frac{r^2+z^2-1}{(r-1)^2 + z^2}\frac{E(m_0)}{K(m_0)}\right).
\end{equation}   
$K(m_0)$ is given by the expression (\ref{eq2.3b})
and  $E(m_0) = \int\limits_0^{\pi/2} \sqrt{1-m_0 \sin^2 \beta}\,d\beta$ \, is the complete elliptical integral of the second kind with the corresponding parameter:
\begin{equation}\label{eq3.4}
 \displaystyle m_0 = \frac{4 r }{(1+  r)^2 + z^2}~.
\end{equation} 
\begin{figure}
\centering
\includegraphics[width = 80mm]{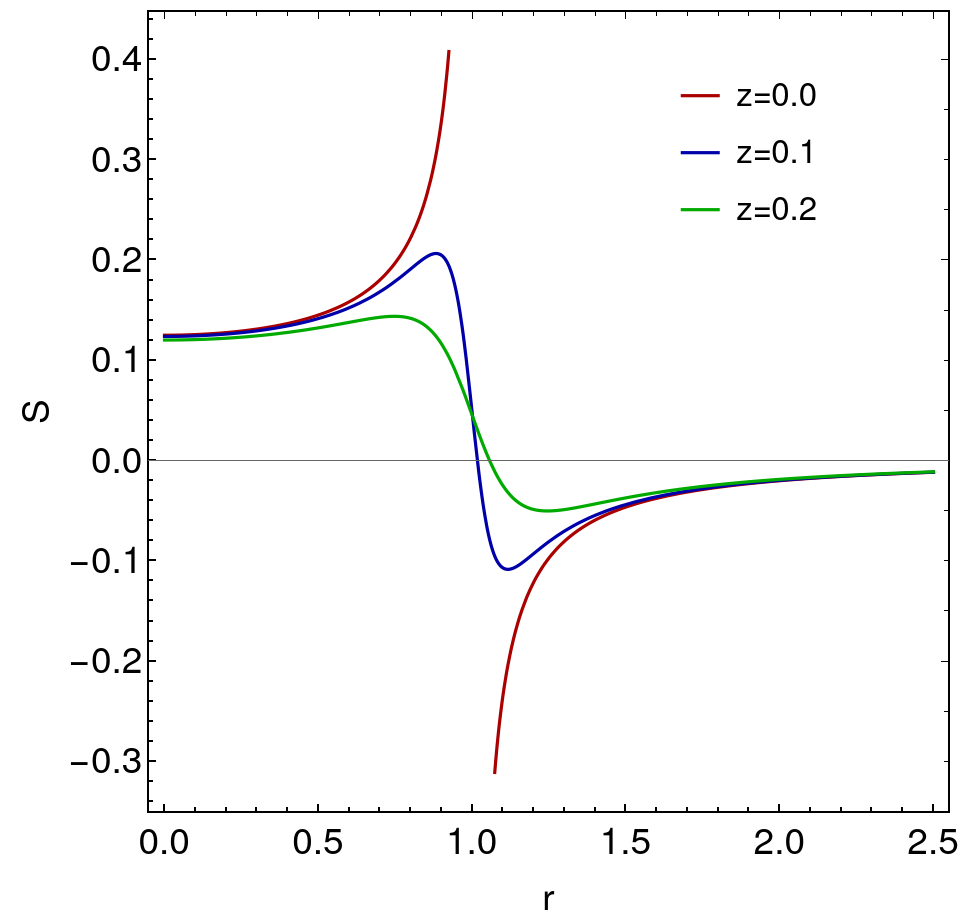}
\caption{Trend of $S$ on the radial coordinate $r$ for three values of the $z$ coordinate.}
\label{fig:S}
\end{figure}
 
\noindent 
The function $S$ is independent of the parameters
of the torus; it only depends on the coordinates of the point. Figure~\ref{fig:S} shows that $S$ reaches the maximal values in the equatorial plane ($z=0$) and in the vicinity of the torus. 
This function corrects the discrepancy between the torus potential and that of the \mc. It reduces the values of the potential provided by the \mc \, within the torus hole ($r < 1 - r_0$) while increasing those on the outer side ($r > 1 + r_0$).
Since $S$ in (\ref{eq3.1}) is { multiplied by} the geometrical parameter $r_0^2$, 
the approximation is better for thinner tori and becomes exact in the limit $r_0 \rightarrow 0$.
We will demonstrate later that a similar mathematical structure can also be used to approximate the potential of a torus with a inhomogeneous density distribution.

From here on out, the expression (\ref{eq3.1}) will be referred to as the S-approximation. 
It provides relative errors that are smaller than those provided by the toroidal series \cite{1973AnPhy..77..279W, Majic2020}. We prove this statement in Appendix~\ref{secA1}, where we compared the two approximations with the exact expression of the potential (\ref{eq2.2}). 
The S-approximation has smaller relative errors, justifying its use for the following purposes.

We initially attempted to apply the same procedure as for the circular case (Maclaurin series expansion in the vicinity of the central \mc), but the approximated expression for the potential turned out to be { rather complicated} (see Appendix~\ref{secA2}).
So we evaluated the possibility of representing the outer potential of a homogeneous torus with an elliptical cross-section by the sum of the potentials of two \mcs. As we will show below, this approximation works well and  enables us to uncover new insights into the gravitational potential of the torus. 

We were inspired by our previous treatment of the circular case and the obvious symmetry of the prolate case, where the two \mcs \, must have the same mass and be placed at equal distances $l$ from the equatorial plane along the major axis of the elliptical cross-section. We decided to test this same configuration for the oblate case as well. 
We therefore imposed the following conditions on a model which has a high degree of symmetry and holds for both the oblate ($\alpha \leq 1$) and the prolate ($\alpha > 1$) case:
\begin{itemize}
\item[$-$] 
    The two \mcs \, are placed symmetrically along the major axis at equal distances, $l$, from the center of the cross-section (Fig.~\ref{fig:scheme}; {\it bottom}).  
{ Numerical experiments (see Appendix~\ref{secA2add}) show that $l$ must be chosen to be proportional to the focal length} \mbox{$f = r_0 \sqrt{1 - \alpha^2}$}: 
    \begin{equation}\label{eq3.4bis}
    l=k f= kr_0 \sqrt{1 - \alpha^2}, 
    \end{equation}
where $0\le k = \text{const}<1$. This choice also satisfies the limiting case for the circular torus ($\alpha = 1$).
    \item[$-$] The mass of each of the two \mcs \, is half the mass of the torus. 
    \item[$-$] The approximate expression for the outer potential of an elliptical torus must coincide with the expression (\ref{eq3.1}) in the limiting case of the circular cross-section. To meet this requirement, we simply modified the prefactor of $S$ in (\ref{eq3.1}) from $r_0$ to $\alpha r_0$.
    \end{itemize}
\noindent
As a result, the expression approximating the outer gravitational potential of a homogeneous torus with an elliptical cross-section takes the form:
\begin{equation}\label{eq3.5}
\varphi^\text{app}_\text{tor} (r,z)  = 
             \Big(\potmc(r,z;-l)+\potmc(r,z;+l)\Big)
            \Big(1 - \alpha^2r_0^2 S(r,z)\Big),
\end{equation}
hereafter called S$_\alpha$-approximation.
The potentials of the two \mcs \, are given by (\ref{eq2.3}) for $x'=\pm l$ and $z'=0$ and are equal to
\begin{equation}\label{eq3.6}
 \potmc(r,z;\pm l) = \frac{G M}{2\pi R}\frac{2(1 \pm l)}{\sqrt{(1\pm l+r)^2 + z^2}}  \, K(m_\pm),
\end{equation}
with the parameters of the elliptical integral:
\begin{equation}\label{eq3.7a}
  \displaystyle m_\pm = \frac{4 r \, (1 \pm l)}{(1 \pm l+ r)^2 + z^2}~.
\end{equation}  

An unknown quantity in the expressions (\ref{eq3.5}) -- (\ref{eq3.7a}) is the constant $k$ which gives the distance of each of the \mcs \, from the cross-section center (\ref{eq3.4bis}) in units of the focal length. In order to determine its value, we minimized the residual differences between the potential calculated from the approximate expression (\ref{eq3.5}) and the exact formula (\ref{eq2.2})
through a set of numerical experiments. We fixed $r_0$  
and varied $\alpha$ and $k$ to obtain maps of the mean absolute errors (MAE) for a grid of points with coordinates ($r_i, z_j$); \mbox{$i,j=1 \dots N$}: 
\begin{equation}\label{3.7b}
\text{MAE}(r_0,\alpha, k) = \frac{1}{N^{\prime}}\sum_{i=1}^{N}{\vphantom{\sum}}^\prime\sum_{j=1}^{N}{\vphantom{\sum}}^\prime |\text{RE} (r_i, z_j; r_0,\alpha, k)|.
\end{equation}
$\text{RE}$ is the relative error in each point given by
\begin{equation}\label{eq3.7c}
\text{RE}(r_i, z_j; k) = 100\kolr{\times} \Big(1- \frac{\varphi^\text{app}_\text{tor} (r_i, z_j;  k)}{\varphi_\text{tor} (r_i, z_j)}\Big).
\end{equation}
The prime attached to the summations in (\ref{3.7b}) indicates that we took into account only those $N'$ points outside the torus volume.

Our numerical experiments for the oblate configuration covered the intervals \mbox{$k= 0.2,..,0.8$} and \mbox{$\alpha= 0.4,..,0.9$} for a set of values of $r_0$ in the range $[0,0.7]$. We calculated the errors in a $N\times N$ grid of discrete points within a rectangular region symmetric with respect to the $r$-axis and with one side coincident with the $z$-axis.  We selected \mbox{$N = 120$} and the steps $\Delta r = \kolg{0.015}$ and $\Delta z= \kolg{0.017}$. 
The calculations made with wider areas proved that this choice included the whole region where the deviations between the exact formula and the model are more significant. Similar numerical experiments were made for the prolate case.

\begin{figure}
\centering
\includegraphics[width = 66mm]{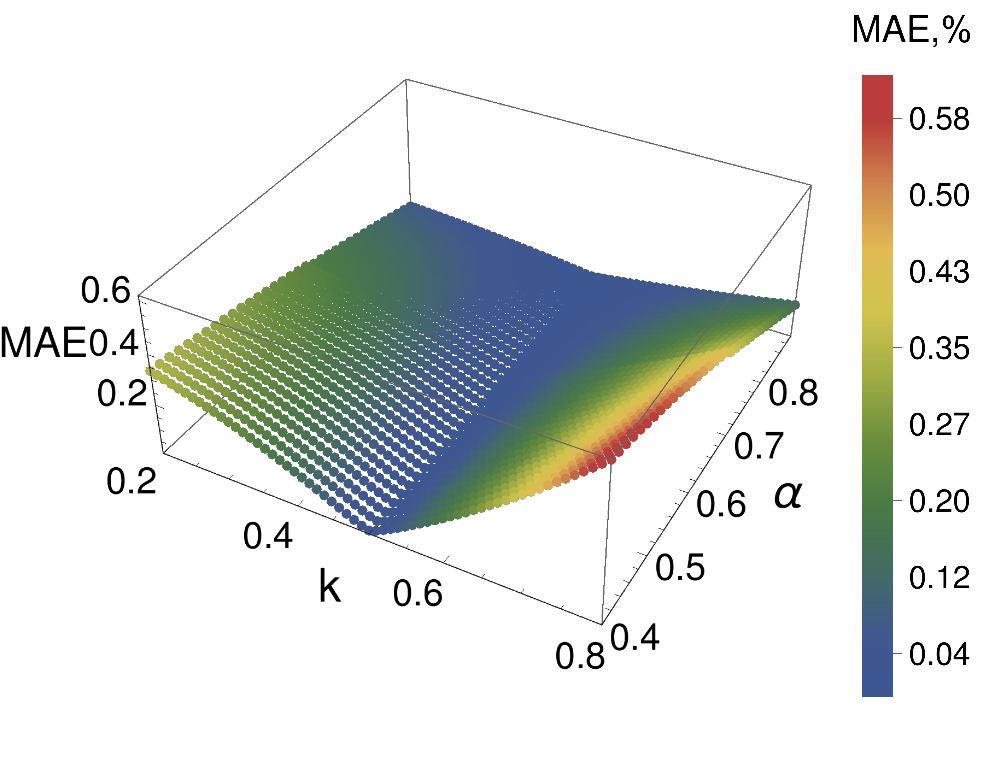}\qquad
\includegraphics[width = 53mm]{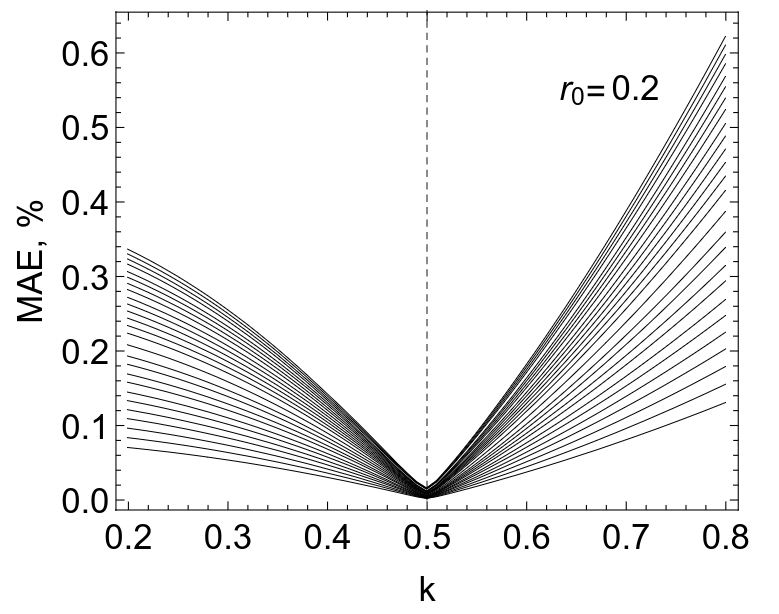}
\caption{Example of the dependence of the mean absolute error, MAE (see expression (\ref{3.7b})), on the axis ratio ($\alpha$) and the distance of the each of the \mcs \, from the center of the oblate cross-section (parameter $k$). The torus parameters are $r_0 = 0.2$ and $\alpha = [0.4,...,0.9]$. The sampling steps are $\Delta k =0.015$, $\Delta \alpha = 0.01$. {\it Left panel}: 
two-dimensional map.  {\it Right panel}: projection onto the MAE-$k$ plane. It is apparent that a minimum occurs at $k=0.5$, regardless of the flattening $\alpha$. We have adopted $M=R=G=1$ for the unit system in this and all other similar figures.}
\label{fig:MAE}
\end{figure} 

Figure~\ref{fig:MAE} shows an example of the error map for $r_0=0.2$, where it is apparent that the minimum occurs systematically at $k=1/2$ for all values of axis ratios $\alpha$ (see also Fig.~\ref{fig:scheme} for a scheme of the \mc \, positions in the torus cross-section).
The result holds for both the oblate and the prolate configuration.

\begin{figure}[h!]
\centering
\includegraphics[width=58mm]{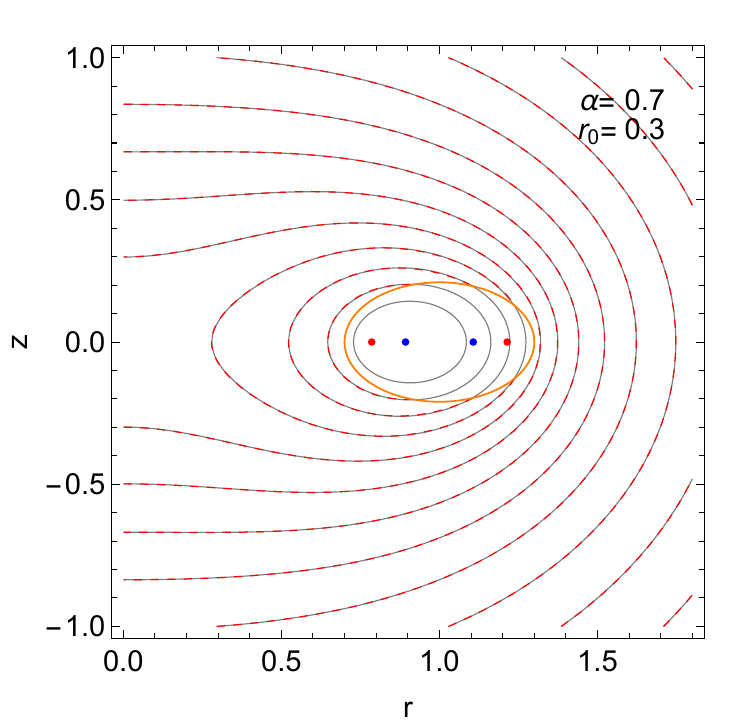}\quad
\includegraphics[width=67mm]{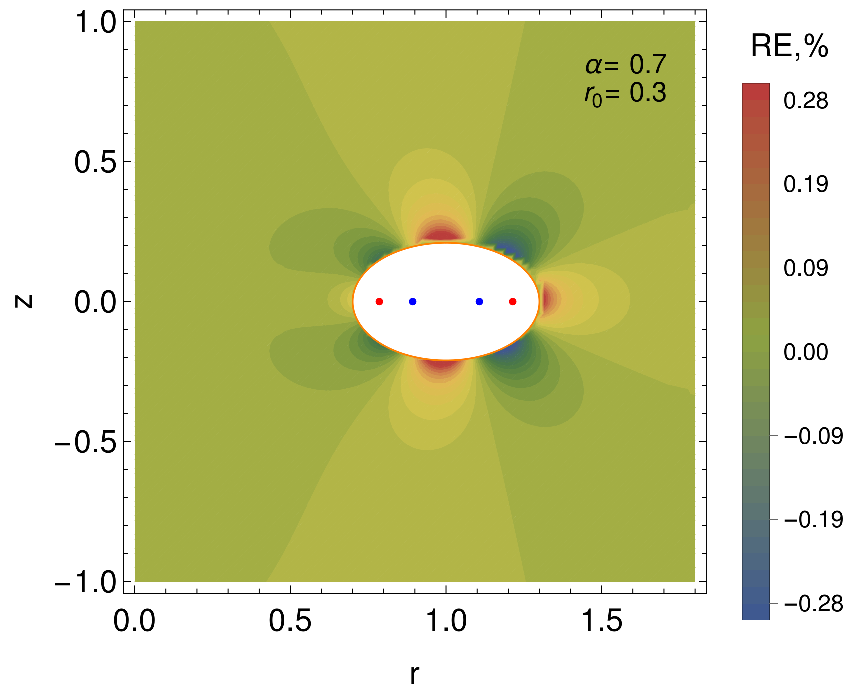}
\includegraphics[width=58mm]{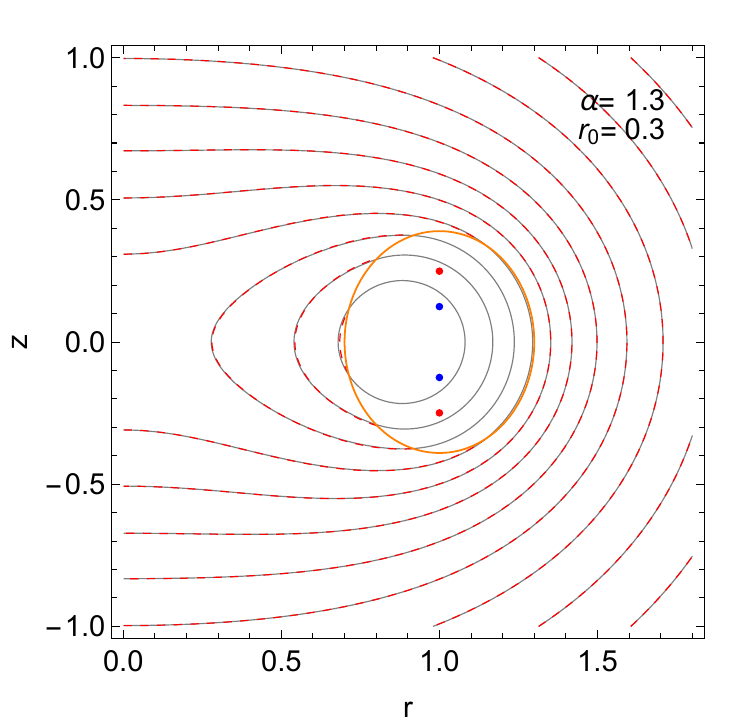}\quad
\includegraphics[width=67mm]{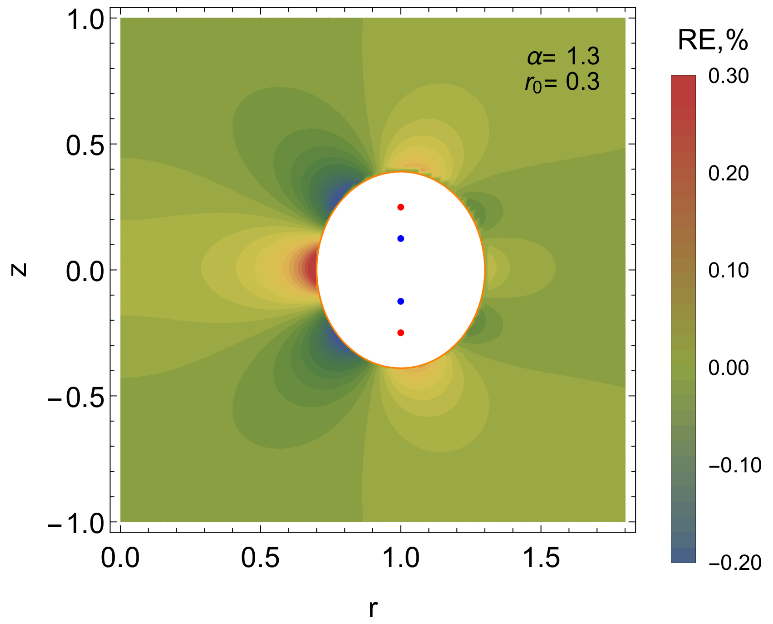}
\caption{{\it Left panel}: level-curves of the outer potential for the oblate ({\it top panel}) and prolate ({\it bottom panel}) configurations, calculated using exact expression (\ref{eq2.2}) (black) and the S$_{\alpha}$-approximation (\ref{eq3.5}) (red dashed). The red dots mark the foci, the blue dots mark the positions of the \mcs.
{\it Right panel}: the corresponding error maps calculated by the formula (\ref{eq3.7c}) for $k=1/2$.}
\label{fig:contour-err}
\end{figure}

In conclusion, we have found that, within $0.6\le\alpha \le 1$ and $0<r_0 \le 0.7$ for the oblate case and within $1\le\alpha < 1.4$ and $0<r_0\le 0.7$ for the prolate case, $l=f/2$ is the value securing the best match of equation (\ref{eq3.5}) to the exact expression (\ref{eq2.2}) of the gravitational potential of a homogeneous torus with an elliptical cross section. Within the quoted ranges for the torus parameters, the errors do not exceed 1\%.
Examples of the isopotential curves with corresponding error maps are shown in 
Fig.~\ref{fig:contour-err} for the oblate ({\it top panels}) and prolate ({\it bottom panels}) cross-sections.  
 
We can therefore formulate the following statement which applies equally well to both oblate and prolate cases.\\

\noindent 
\textit{Statement 1. The outer potential of a homogeneous torus with an elliptical cross-section can be effectively represented (with errors smaller than 1\%) by the S$_\alpha$-approximation (\ref{eq3.5}) with two \mcs \,  located symmetrically along the major axis of the cross-section at half of the distances to the foci.}\\

\noindent
The validity of this result was \kolr{verified} by extensive numerical experiments with different combinations of distances ($l_1$ and $l_2$). Examples of them are reported in Appendix~\ref{secA2add}.
 
While we  have not yet found a physical justification to this conjecture,  it probably provides an additional clue to understanding the gravitational properties of the torus potential. The two \mcs \, approximation (\ref{eq3.5}) with ($l=f/2$) matches the exact expression with smaller relative errors
than the Maclaurin series expansion at the same order of approximation in the vicinity of the central \mc\ (see Appendix~\ref{secA2} and Fig.~\ref{fig:Maclaurin-twor-comp}). As we will see below (Section~\ref{section4}), this representation allows us to find a way to approximate the outer potential even for inhomogeneous density distributions. It should be noted that in the case of prolate tori, the coordinates of the \mcs \, are complex conjugate numbers. This is because the cross-section is projected onto the complex plane and thus the coordinates of the foci become complex conjugate (see Appendix~\ref{secA3}).

\subsubsection{Potentials of confocal homogeneous tori}
\label{section3b}
The result of the previous Section has an application to the case of confocal tori, i.e., tori whose elliptical cross-sections have equal focal lengths. 
Consider two such tori with the same total mass $M$, the same major radius $R$, and the same focal lengths $f$, but different flattening $\alpha$.  
We denote the semi-axes of the cross-section of the first torus as $a=r_0$ and $b=\alpha a$, and those of the second torus as $a'$ and $b'= \alpha' a'$. 
The approximate expression of the outer potential for the first torus is given by (\ref{eq3.5}). It is readily seen that the potential of the second torus, confocal to the first one, can be obtained by the following substitution in (\ref{eq3.5}): $\alpha^2 r_0^2 \rightarrow \alpha'^2 a'^2$, that is, $b^2 \rightarrow b'^2$. Thus, within the S$_{\alpha}$-approximation, the ratio of the outer potentials of two confocal tori of equal mass and radius is
\begin{equation}\label{eq3.7}
\frac{{\varphi'}^\text{app}_\text{tor}}{\varphi^\text{app}_\text{tor}} = \frac{1-b'\, S(r,z)}{1-b\,S(r,z)}.
\end{equation}
Since  $\mid S \mid <1$ and $b$, $b'<1$, the two confocal tori have the same potential within the limits given by the function $S$. 
It follows that:\\

\noindent 
\textit{Statement 2. At a first  approximation, the ratio of the potentials of two confocal tori with \kolr{ the same \kolb{major} radius and} different masses ($M$, $M'$) is 
\begin{equation}\label{eq3.7add}
\frac{{\varphi'}^\text{\rm \rm app}_\text{\rm tor}}{\varphi^\text{\rm app}_\text{\rm tor}} = \frac{M'}{M}.
\end{equation}
}\\
This result has some analogy with the Laplace theorem on the outer potential of confocal ellipsoids. 

\section{\kolr{ Torus with an elliptical cross-section and an inhomogeneous density distribution}}
\label{section4}
\subsection{The potential for isodensity contours of constant axis ratios} 
\label{subsection4.1}
\begin{figure}[h!]
\centering
\includegraphics[width = 60mm]{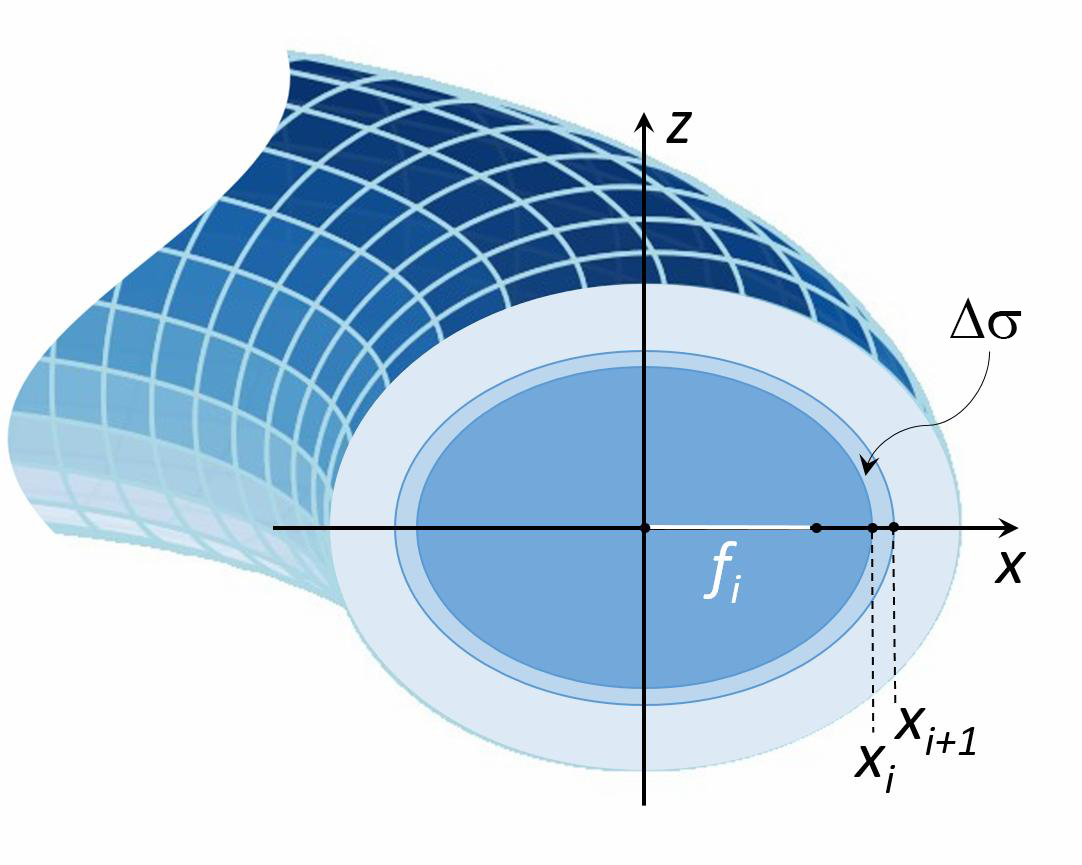}
\caption{Sketch of a torus with self-similar isodensity contours.}
\label{fig:inhomscheme}
\end{figure}

The above results allow us to investigate the properties of the outer potential of an inhomogeneous torus with elliptical density contours. We can represent it as the sum of nested homogeneous tori with different sizes of the elliptical cross-sections, $R_0$, but the same radius $R$ and flattening/elongation $\alpha$ { (Fig. \ref{fig:inhomscheme})}. 

Taking into account { the expression} (\ref{eq3.5}), we can write that 
\begin{equation}\label{eq4.1}
\varphi_\text{tor} (r,z) \approx 
             \sum\limits_i \bigtriangleup\den(x_i)\, \varphi^\text{app}_\text{tor}(r,z;l_i)\, A(x_i),
\end{equation} 
where $x$ is the coordinate counted from the center of the torus cross-section ($x=r-1$), $\bigtriangleup \den_i$ is the surface density per unit mass of a shell (between $x_i$ and $x_{i+1}$), $A(x_i) = \pi \alpha x_i^2$ is the area of the  cross-section, { \kolr{and} $l_i = f_i/2 = x_i\sqrt{1-\alpha^2}/2$} is the distance from the center of the pair of \mcs \,  approximating the potential of the $i$-th nested homogeneous torus whose cross-section major semi-axis is $x_i$.

Moving from the sum to the integration, we obtain the expression for the outer potential of the inhomogeneous elliptical torus:
\begin{equation}\label{eq4.2}
\varphi_\text{tor}(r,z)  \approx 
             \int\limits_0^{r_0} A(x) \frac{\partial\den(x)}{\partial x}\, \varphi^\text{app}_\text{tor}(r,z;x)\,dx,
\end{equation} 
where $\den(x) = \den_0 \, \kolr{h(x)}$ is the surface density of the torus cross-section per unit mass with the normalization factor 
\begin{equation}\label{eq4.3}
\displaystyle \den_0 = \left[{\int\limits_0^{r_0} A(x)\displaystyle{\frac{\partial \kolr{h(x)}}{\partial x}}dx}\right]^{-1},
\end{equation} 
 and $\varphi^\text{app}_\text{tor}(r,z;x)$ is given by (\ref{eq3.5}) with the substitution: \mbox{$r_0 \rightarrow x$}. 
The formula (\ref{eq4.2}) is valid for any radial density distribution with elliptical iso-contours. 

\subsection{Confocal density distribution}
\label{subsection4.2}
\begin{figure}[h!]
\centering
\includegraphics[width = 65mm]{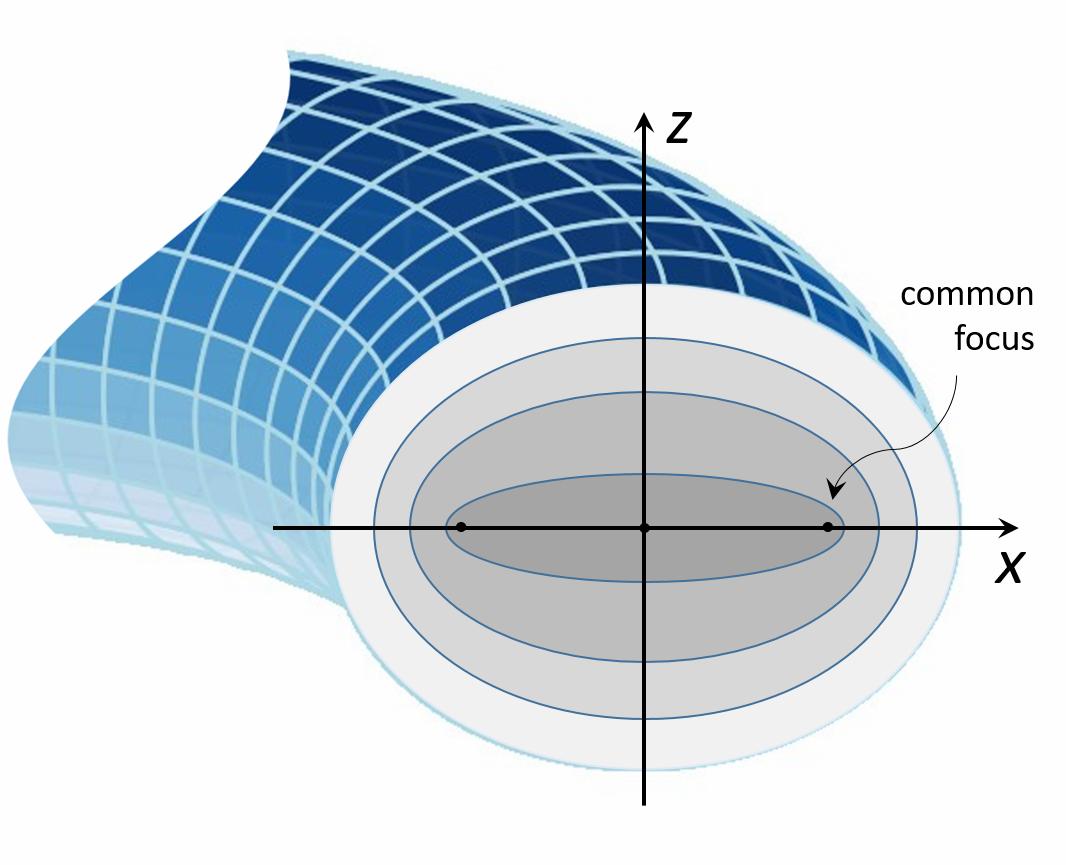}
\caption{\kolr{ Sketch of a torus with confocal isodensity contours.}}
\label{FigConfocalTorus}
\end{figure}

We  now consider a special case of a torus with a cross-section characterized by isodensity contours which are confocal ellipses. The corresponding density distribution is one step more realistic than a plain homogeneous distribution as it smooths the abrupt discontinuity present at the boundary  of the torus body (see Fig.~\ref{FigConfocalTorus}). 

\subsubsection{Analytical approach} \label{subsubsection4.2}
As above, we replace the torus with the sum of nested homogeneous tori with different sizes of the elliptical cross-sections, but all with the same focal distances.
This implies that $l=f/2=$ const for all the nested confocal tori.
Therefore, from { the expression} (\ref{eq4.2}), using $f$ as the lower limit in the integral\footnote{For the particular case of a confocal density distribution, the surface density along the line connecting the two foci is constant, that is: ${d\kolr{h(x)}}/{dx}=0$ in the interval $[0,f]$.
It means that the lower limit in the expression (\ref{eq4.3}) can be set to $f$
as the integral is zero up to this value.}, we have:
\begin{equation}\label{eq4.4}
\varphi_\text{tor} (r,z)\approx \Big( \potmc(r,z;-l)+\potmc(r,z;+l)\Big) \,\den_0
        \times \int\limits_f^{r_0}  \frac{\partial \kolr{h(x)}}{\partial x}A(x)(1 - F(r,z;x))dx\, , 
\end{equation}
\kolr{where $F(r,z;x)=(x^2 - f^2) S(r,z)$.}
\kolr{ We have exploited the fact that $\varphi^\text{app}_\text{tor}$, determined by the expression (\ref{eq3.5}), can be taken out of the integral because in a confocal configuration $l$ does not change: $\varphi^\text{app}_\text{tor}=\varphi^\text{app}_\text{tor}(r,z;l)=\varphi^\text{app}_\text{tor}(r,z;f,r_0)$.}
After some \kolr{simplifications, using} (\ref{eq4.3}) we obtain:
\begin{equation}\label{eq4.4a}
\varphi_\text{tor} (r,z) \approx 
           \Big( \potmc(r,z;-l)+\potmc(r,z;+l)\Big) \left[1 - \left(\den_0 \, H - f^2\right)S(r,z) \right],  
\end{equation} 
where
\begin{equation}\label{eq4.4b}
        H = \int\limits_f^{r_0} x^2 \frac{\partial \kolr{h(x)}}{\partial x}A(x)dx.  
\end{equation} 
Since the functions \kolr{$h(x)$} and $A(x)$ are continuous in the interval $[f,r_0]$, we can apply the generalized mean-value theorem, \kolr{ with which} the expression (\ref{eq4.4b}) takes the form: 
\begin{equation}\label{eq4.4d}
        H \approx \frac{r_0^2+f^2}{2}\int\limits_f^{r_0} \frac{\partial \kolr{h(x)}}{\partial x}A(x)dx.  
\end{equation} 
The final approximate expression of the outer potential of the torus with a confocal density distribution is
\begin{equation}\label{eq4.6}
\displaystyle
\varphi_\text{tor} (r,z)  \approx 
           \left( \potmc(r,z;-l)+\potmc(r,z;+l)\right) \left(1-\frac{r_0^2-f^2}{2}S(r,z)\right).
\end{equation} 
The factor with the density distribution disappears because \kolr{ $l =\text{const}$}. In conclusion, we can \kolr{ state the following}. \\

\noindent\textit{Statement 3. The outer potential of a torus with a confocal density distribution in its cross-section is only weakly dependent on the density distribution law.}\\

\noindent 
We will now show an application to a Gaussian density distribution for both flattened/elongated tori.

\begin{figure}[h]
\begin{center}
\begin{minipage}[h]{0.40\linewidth}
\begin{center}
\includegraphics[width=\linewidth]{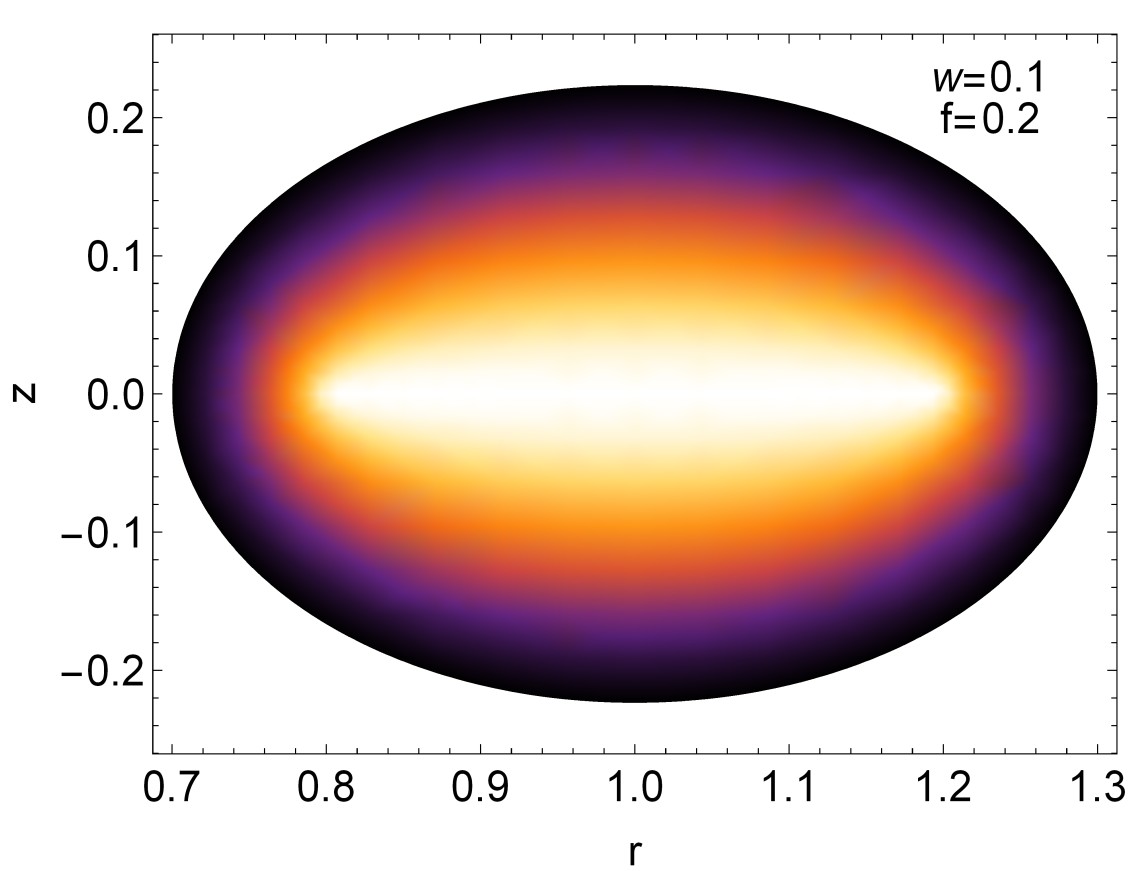} 
\end{center} 
\end{minipage}
\hspace{5mm}
\vspace{0.01 cm}
\begin{minipage}[h]{0.40\linewidth}
\begin{center}
\includegraphics[width=1\linewidth]{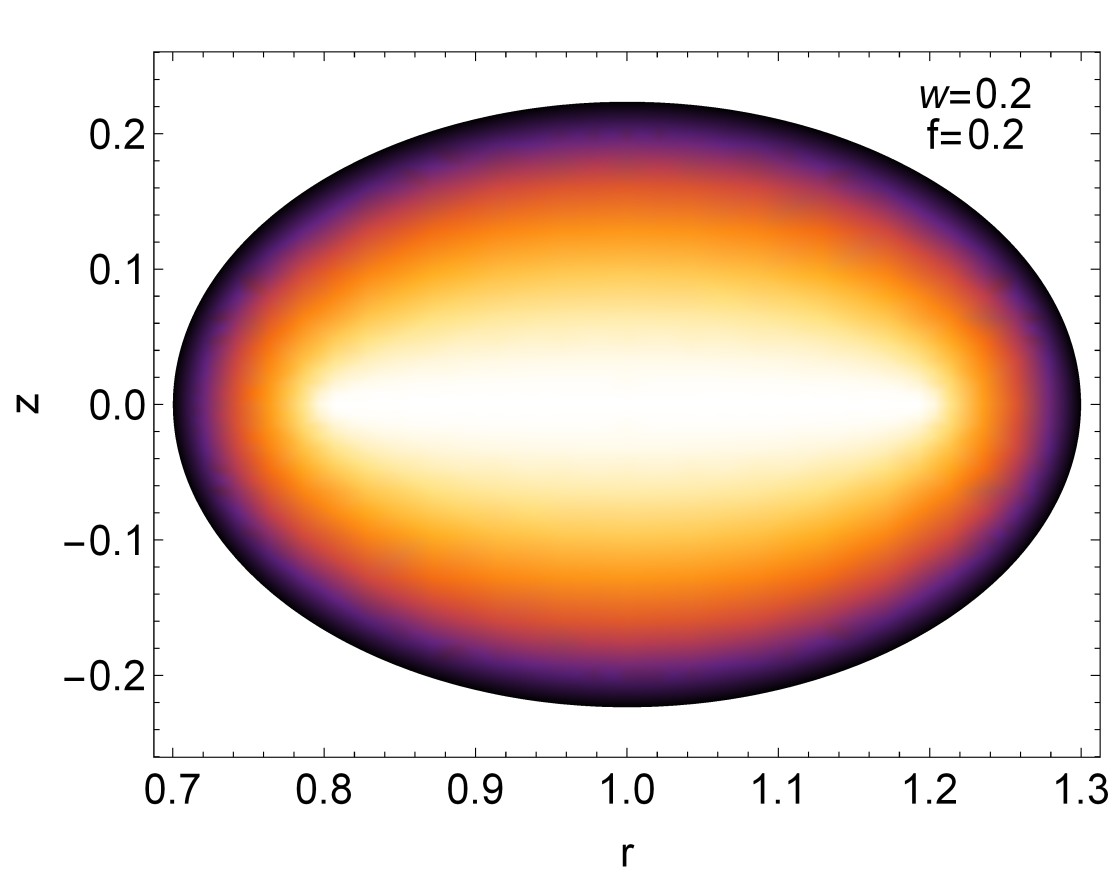} 
\end{center}
\end{minipage}

\begin{minipage}[h]{0.40\linewidth}
\begin{center}
\includegraphics[width=1\linewidth]{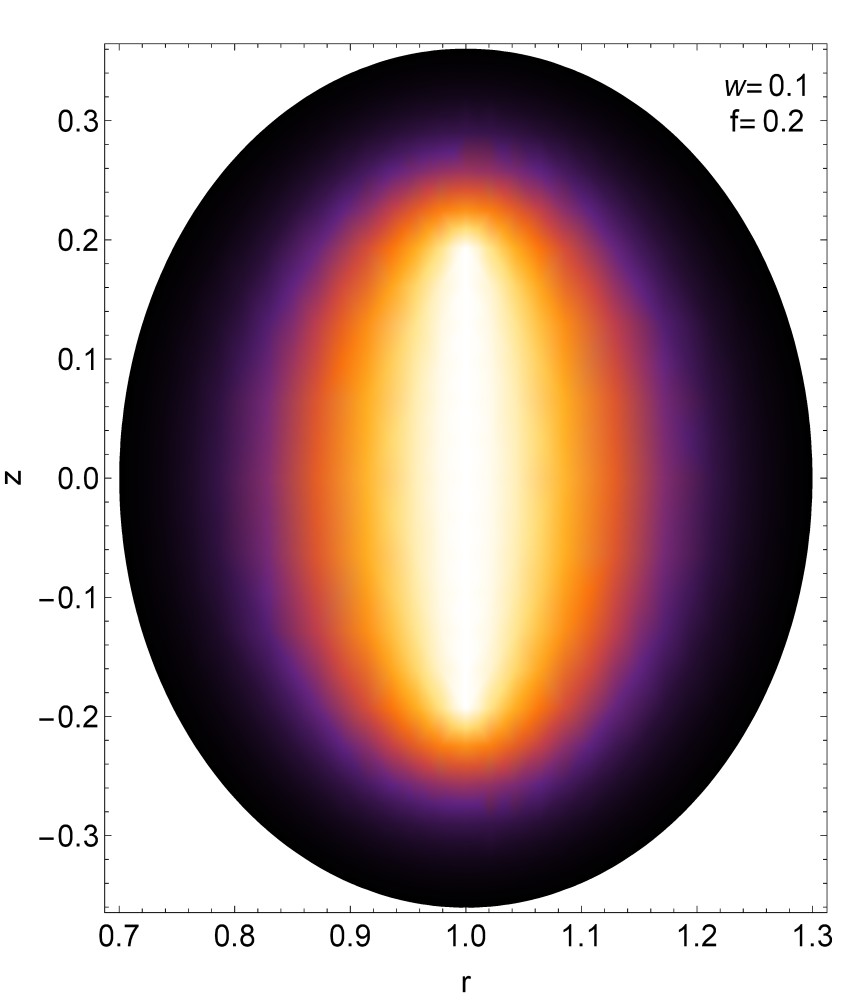}
\end{center}
\end{minipage}
\hspace{5mm}
\begin{minipage}[h]{0.40\linewidth}
\begin{center}
\includegraphics[width=1\linewidth]{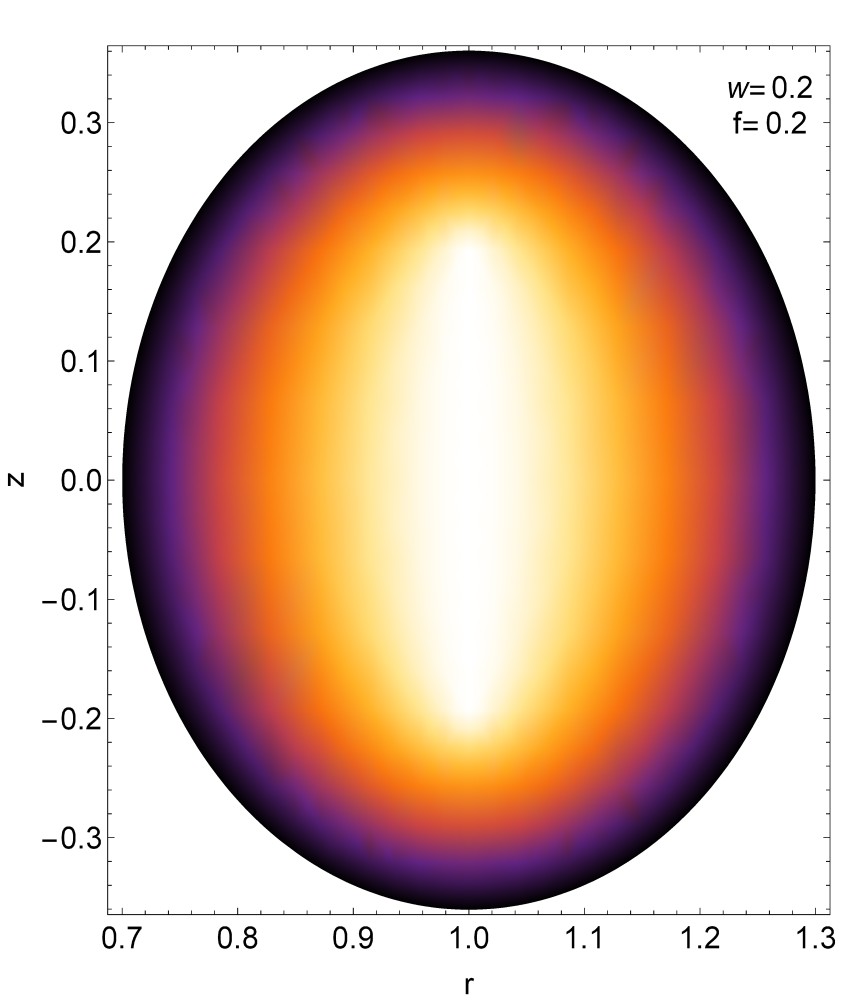}
\end{center}
\end{minipage}
\caption{Density distribution color maps in the elliptical cross-section of a torus with confocal isodensity contours and a density profile given by the function(\ref{eq4.7}), for $r_0 = 0.3$ and two values of the dispersion, {$w = 0.1$} ({\it left panels}) and {$w = 0.2$} ({\it right panels}). The focal length for both the oblate case  ({\it top panels}) and  the prolate case ({\it bottom panels}) is $f=0.2$.}
\label{fig:condistr}
\end{center}
\end{figure}

\subsubsection{Application to a Gaussian density distribution}\label{subsection4.3}
As an example, let us consider the Gaussian density distribution in the torus cross-section:
\begin{equation}\label{eq4.7}
\displaystyle
\den(x,z) = \den_0 \exp\left[-\frac{a(x,z)^2}{2w^2}\right],
\end{equation} 
where $a$ is the semi major axis of a confocal isodensity contour
and again the coordinate $x$ is counted from the center of the torus cross-section ($x=r-1$); outside the torus boundary $\den=0$.

To obtain the expression for $a$ corresponding to the confocal case, we must solve { equation} (\ref{eq2.1}) with the substitution $r_0=a$ and $\alpha = \sqrt{1-f^2/a^2}$ for the fixed focal length:
\begin{equation}\label{eq4.8}
\displaystyle
a(x,z) = \frac{1}{\sqrt{2}}\left[\xi +\sqrt{\xi^2 - 4f^2x^2}\right]^{1/2},
\end{equation}
where $\xi = x^2+z^2+f^2$.

Figure~\ref{fig:condistr} maps the confocal density distribution in the torus cross-section normalized to $\sigma_0$ for oblate ({\it top \kolr{ panels}}) and prolate ({\it bottom \kolr{ panels}}) cross-sections. 
\begin{figure}[h!]
\centering
\includegraphics[width=51mm]{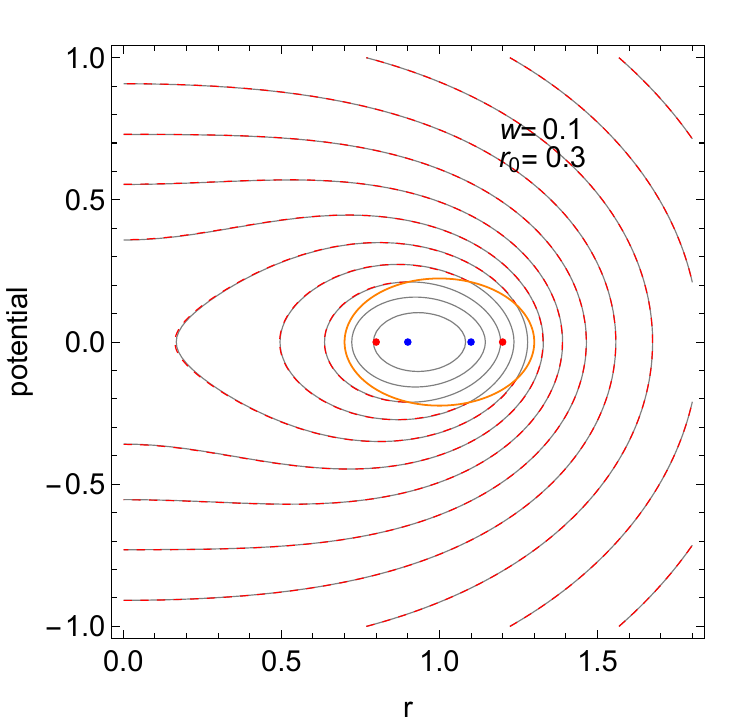}\quad
\includegraphics[width=61mm]{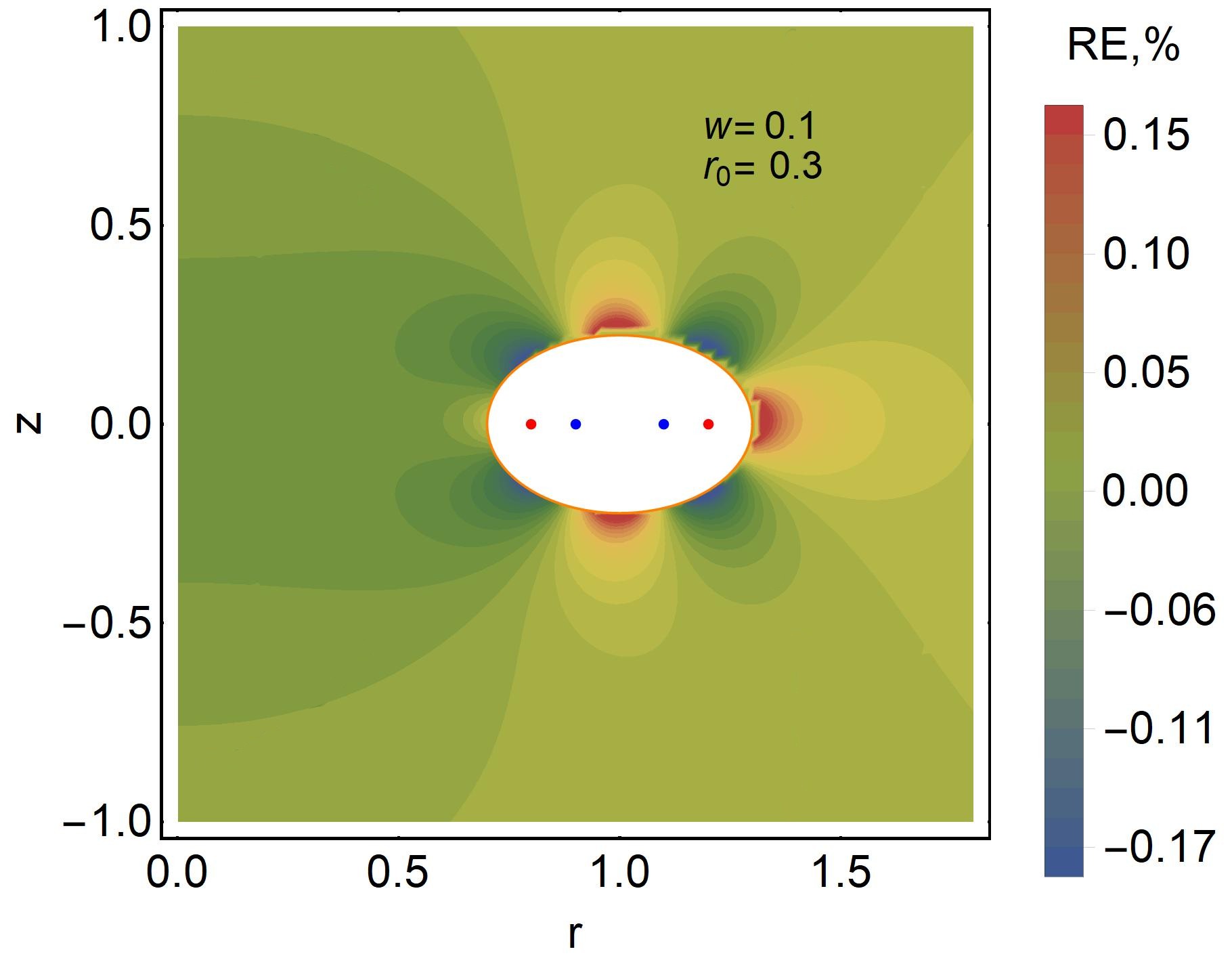}
\includegraphics[width=51mm]{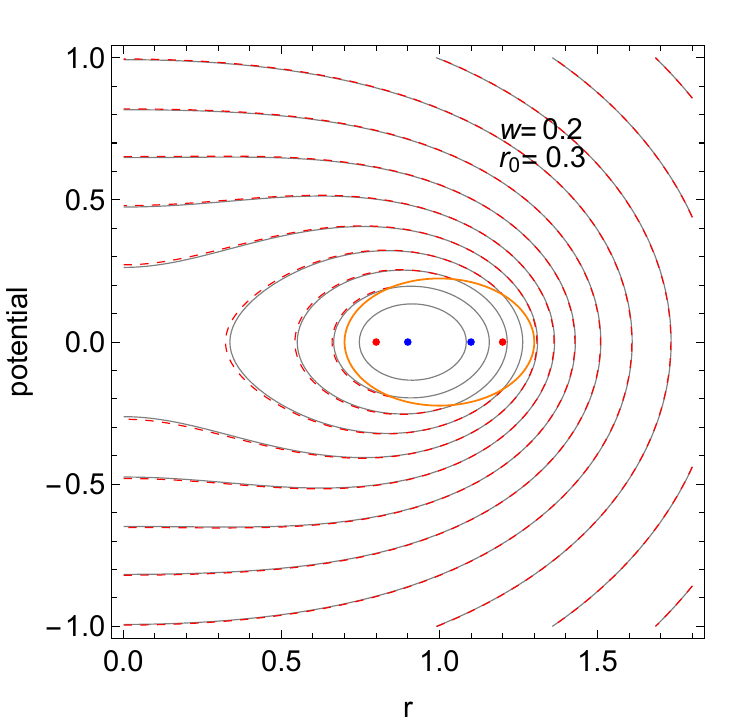}\quad
\includegraphics[width=61mm]{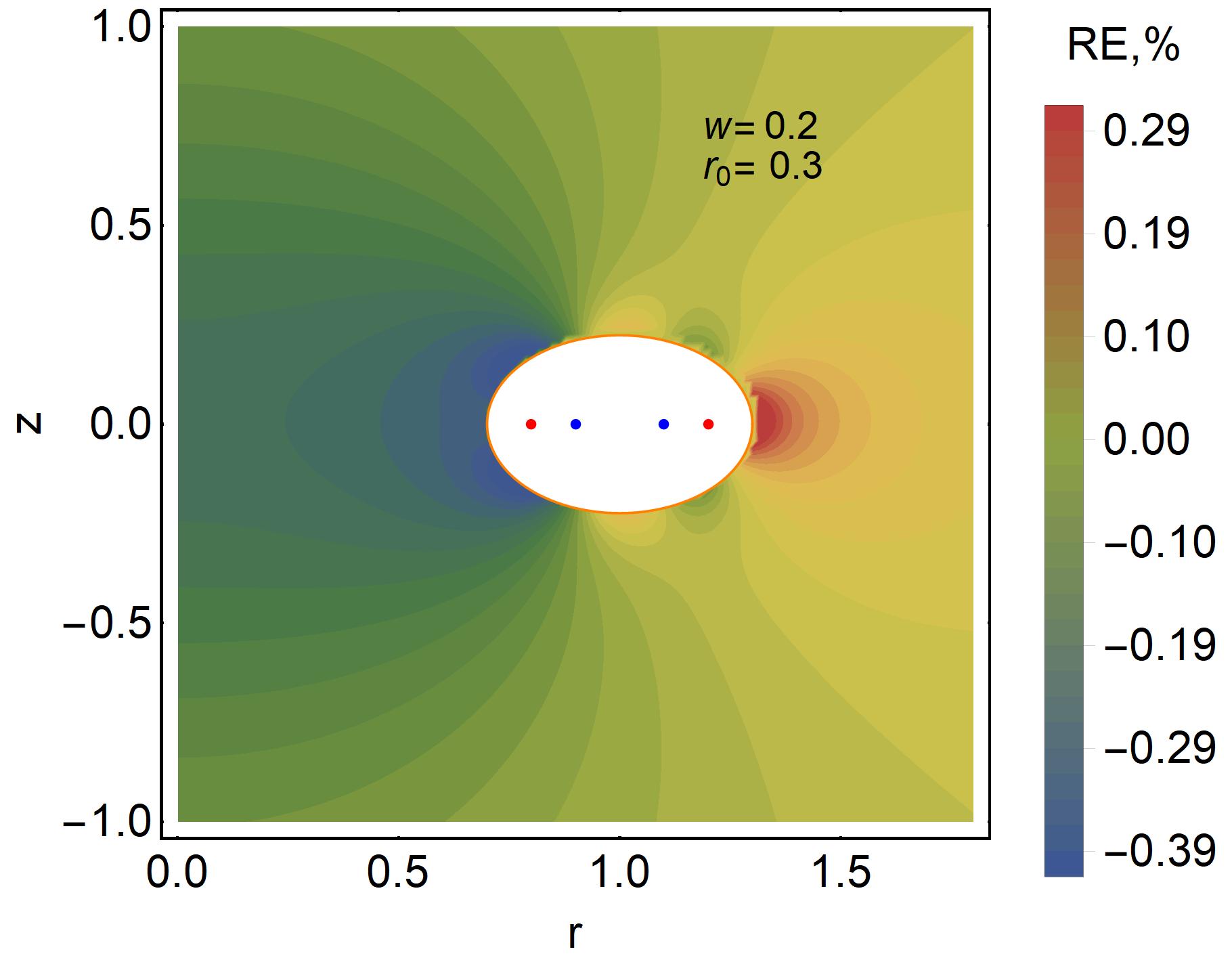}
\caption{{{\it Left panel}: isopotential curves calculated using the exact formula (\ref{eq4.9}) (black) and the approximate expression (\ref{eq4.6}) (red), for an oblate cross-section ($f=0.2$) and two values of the dispersion: $w=0.1$ ({\it top panel}) and $w=0.2$ ({\it left} panels}). 
{\it Right panel}: the relative error maps.}
\label{fig:contour-err-inhom}
\end{figure}

\begin{figure}[h!]
\centering
\includegraphics[width=48mm]{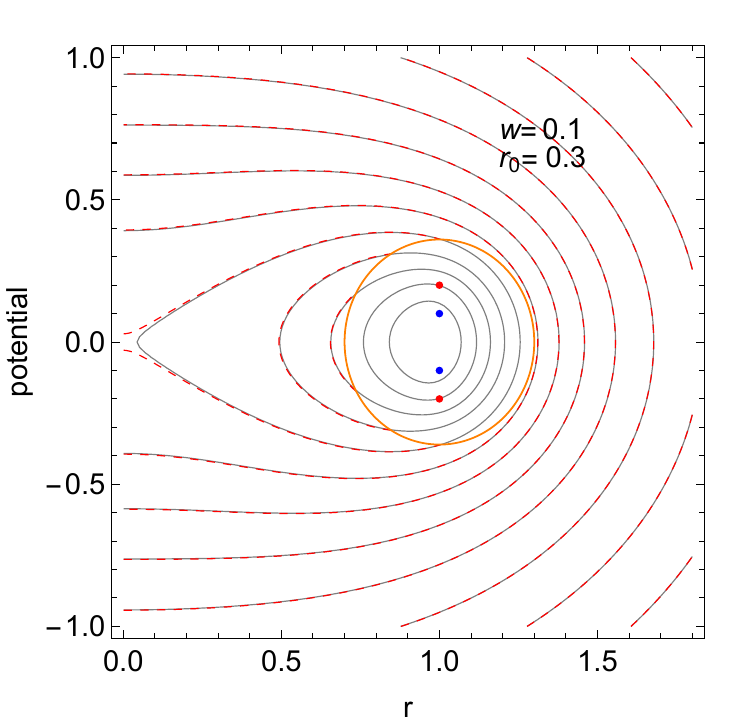}\quad
\includegraphics[width=61mm]{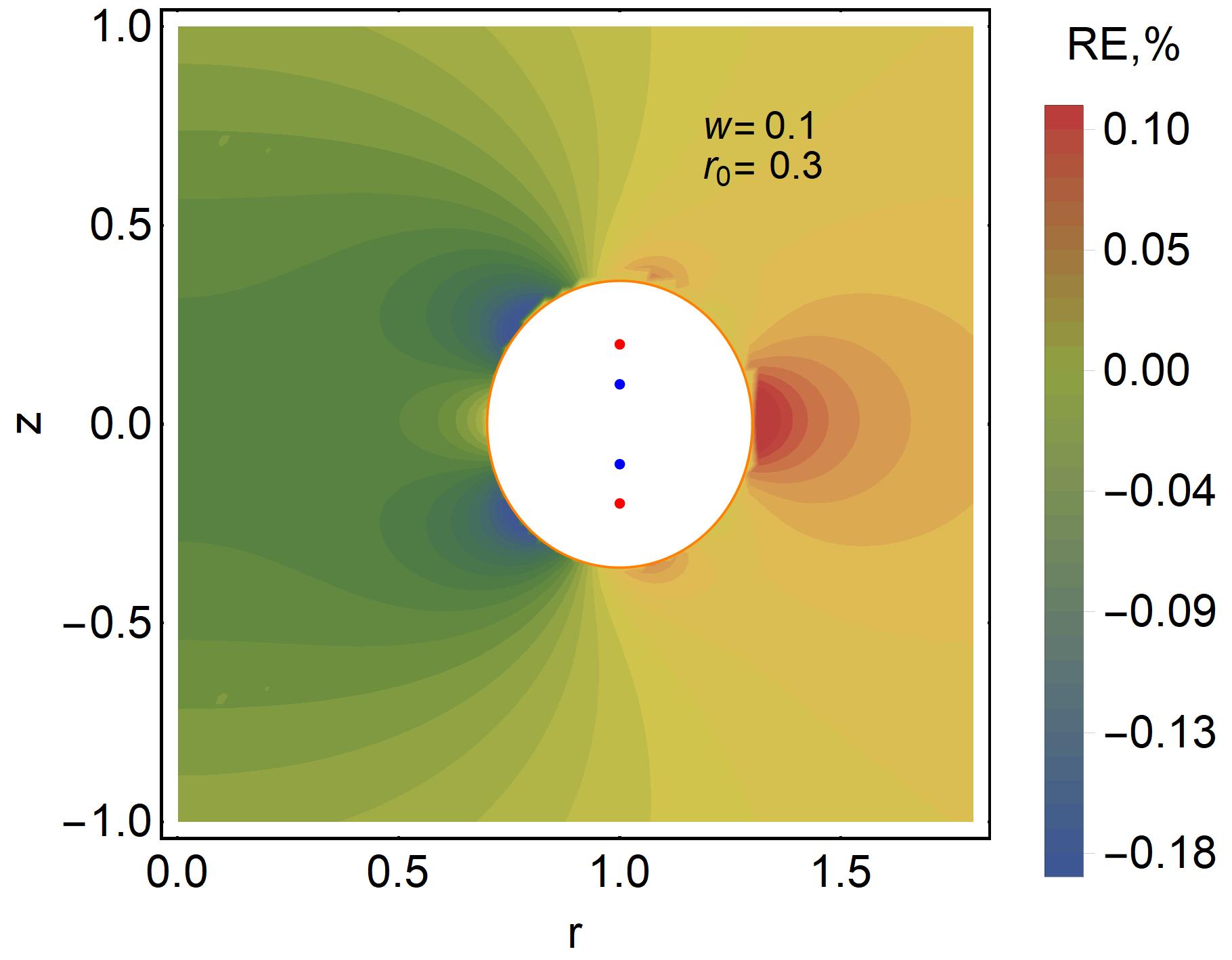}
\includegraphics[width=48mm]{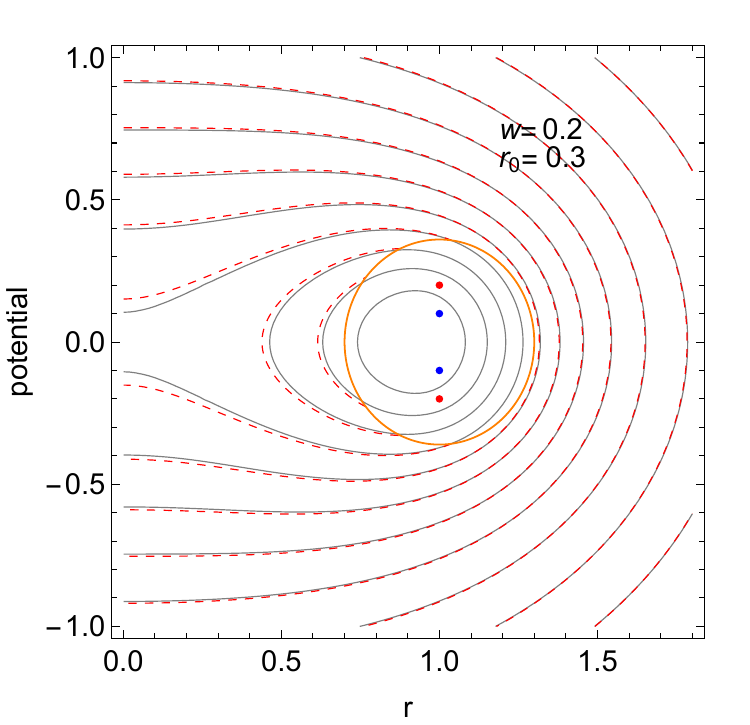}\quad
\includegraphics[width=61mm]{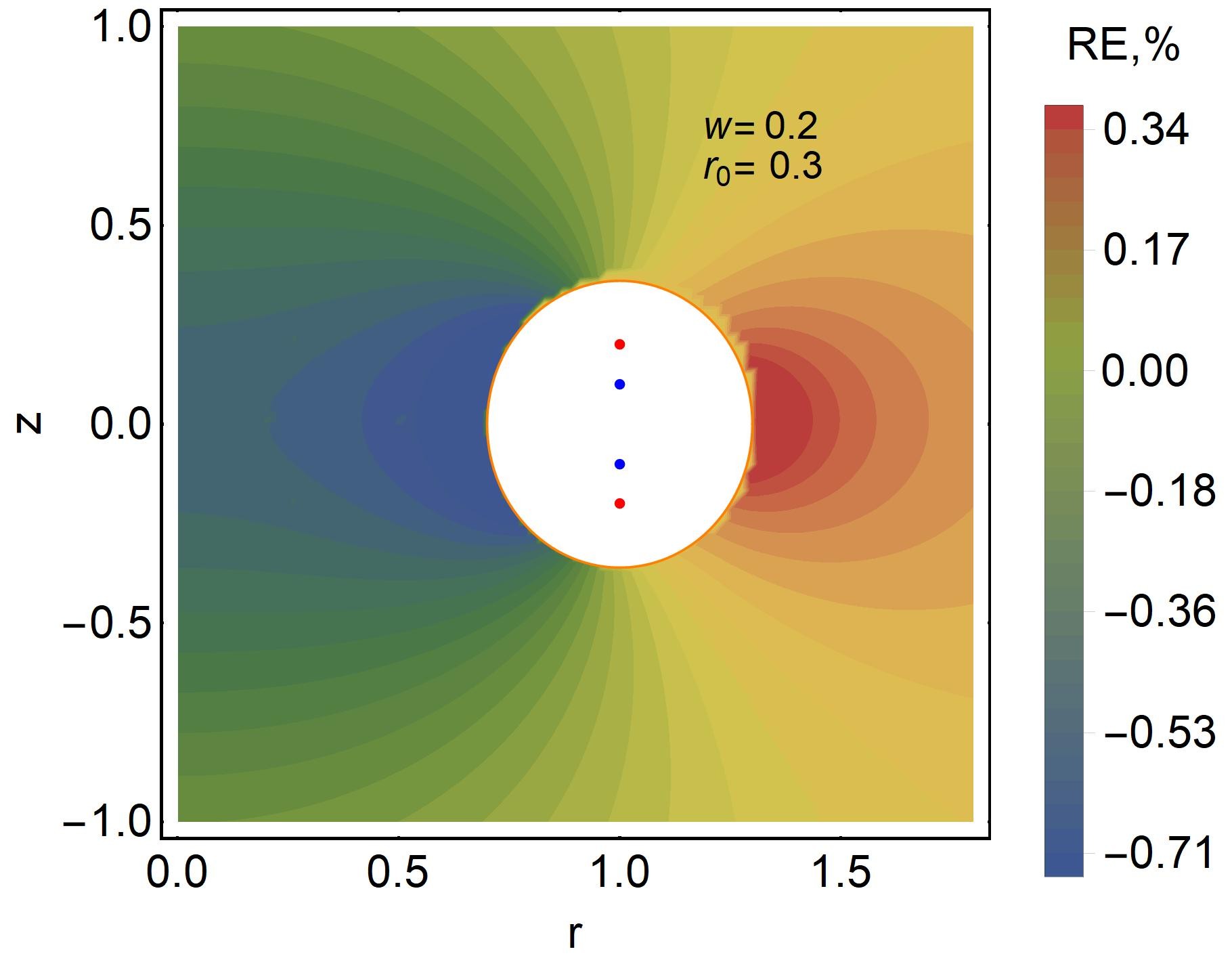}
\caption{{\kolr{ Same as \kolr{ in} Fig.~\ref{fig:contour-err-inhom} but for a} prolate cross-section ($f=0.2$}).}
\label{fig:contour-err-inhom-pr}
\end{figure}

\kolr{ It should be noted} that in the considered prolate case (Fig.~\ref{fig:condistr}, {\it bottom \kolr{ panels}}), the parameter $r_0$ is the same as in the oblate case  and  consequently the area of its cross-section is larger.
Substituting the density law (\ref{eq4.7}) in (\ref{eq2.2}), we obtain the expression for the torus potential:
\begin{equation}\label{eq4.9}
\varphi_\text{tor} (r,z) = \den_0 
	\int\limits_{-r_0}^{r_0} 
        \int\limits_{-z_e}^{z_e}
             \exp\Big[-\frac{a(x',z')^2}{2w^2}\Big]\potmc(r,z; x', z')dz'dx',
\end{equation} 
where
\begin{equation}\label{eq4.10}
\displaystyle \den_0 = \left[{\int\limits_{-r_0}^{r_0} 
        \int\limits_{-z_e}^{z_e} \exp\left[\displaystyle{-\frac{a(x',z')^2}{2w^2}}\right]dz'dx'}\right]^{-1}\, .
\end{equation} 
Equation (\ref{eq4.9}) is exact everywhere inside and outside the torus with a Gaussian confocal density distribution in the elliptical cross-section.

Statement 3 allows us to replace this double-integral expression (\ref{eq4.9}) with the two-\mc \, model (\ref{eq4.6}) which does not involve the density law \kolg{ and simplifies the multiple integration with a combination of just elliptical integrals}. 
More importantly, it proves the following: \\

\noindent 
\textit{Statement 4. Being weakly dependent on the density law, the potential of a non-uniform torus with a confocal density distribution in its elliptical cross-section is well approximated (with less than 1\% error) by the S$_\alpha$-approximation of the uniform case.}\\

The left panels in Figs.~\ref{fig:contour-err-inhom} and \ref{fig:contour-err-inhom-pr} present the isopotential curves for two values of $w$, calculated by exact formula (\ref{eq4.9}) and using the two-\mc \, approximation (\ref{eq4.6}). 
The corresponding RE maps for both the oblate and prolate cross-sections show that the errors are less than 1\% (middle and right panels of Figs.~\ref{fig:contour-err-inhom} and \ref{fig:contour-err-inhom-pr}).

\section{The components of gravitational force}
\label{force}
In this section, we derive the gravitational force for a homogeneous torus with an elliptical cross-section. The radial ($F_{\text{tor},r}$) and vertical ($F_{\text{tor},z}$) components for the exact integral expression of the torus potential (\ref{eq2.2}) are:  
\begin{equation}\label{eq5.3}
F_{\text{tor},r}(r, z) = \frac{1}{R}\frac{\partial \varphi_\text{tor}}{\partial r} = \den_0 
	\int\limits_{-r_0}^{r_0} 
        \int\limits_{-z_e}^{z_e}
             F_{\text{itr},r}(r,z;x',z')dz'dx',
\end{equation}  
\begin{equation}\label{eq5.4}
F_{\text{tor},z}(r, z) = \frac{1}{R}\frac{\partial \varphi_\text{tor}}{\partial z}= \den_0 
	\int\limits_{-r_0}^{r_0} 
        \int\limits_{-z_e}^{z_e}
             F_{\text{itr},z}(r,z;x',z')dz'dx'\,.
\end{equation}  
Here the force components of \mc \, $ F_{\text{itr},r} = (1/R)\partial \potmc / \partial r$ and \mbox{$ F_{\text{itr},z} = (1/R)\partial \potmc / \partial z$} take the following form after simplifications:
 \begin{align}\label{eq5.1}
 \displaystyle 
    F_{\text{itr},r}(r,z;x',z')
    &=-\frac{GM}{\pi R^2}\Big[\frac{1+x'+r}{4\sqrt{1+x'}}\left(\frac{m}{r}\right)^{3/2}K(m)- 
    \nonumber\\
    &-\frac{(1+x')^2-r^2+(z-z')^2}{(1+x'-r)^2+(z-z')^2}\frac{\sqrt{(1+x')m}}{2r^{3/2}}\Big(E(m) - (1-m)K(m)\Big)\Big],
  \end{align}
\begin{equation}\label{eq5.2}
    F_{\text{itr},z}(r,z;x',z') =- \frac{GM}{\pi R^2}\frac{z-z'}{(1+x'-r)^2+(z-z')^2}\sqrt{\frac{(1+x')m}{r}}E(m)\,,
\end{equation}
where the module of the complete elliptical integrals corresponds to the expression (\ref{eq2.4}).
As potential, these expressions work at every point: inside and outside the torus' volume. So, we will call (\ref{eq5.3}) - (\ref{eq5.4}) the "exact"\, expressions.

The approximate expressions for the radial and vertical components of torus' gravitational force are the derivatives of (\ref{eq3.5}) by the corresponding coordinates, which after simplifications give us:
\begin{align}\label{eq5.5}
F^\text{app}_{\text{tor},r} (r,z) & = 
             \Big(F_{\text{itr},r}(r,z;-l)+F_{\text{itr},r}(r,z;+l)\Big)
            \Big(1 - \alpha^2r_0^2 S(r,z)\Big) + \nonumber\\
            &+\alpha^2r_0^2 \Big(\potmc(r,z;-l)+\potmc(r,z;+l)\Big)S'_r(r,z),
\end{align}
\begin{align}\label{eq5.6}
F^\text{app}_{\text{tor},z} (r,z) & = 
             \Big(F_{\text{itr},z}(r,z;-l)+F_{\text{itr},z}(r,z;+l)\Big)
            \Big(1 - \alpha^2r_0^2 S(r,z)\Big)+\nonumber\\
            &+\alpha^2r_0^2 \Big(\potmc(r,z;-l)+\potmc(r,z;+l)\Big)S'_z(r,z),
\end{align}
where 
 \begin{align}\label{eq5.7}
 \displaystyle 
    F_{\text{itr},r}(r,z;\pm l)
    &=-\frac{GM}{2\pi R^2}\Big[\frac{1 \pm l+r}{4\sqrt{1\pm l}}\left(\frac{m_\pm}{r}\right)^{3/2}K(m_\pm)- 
    \nonumber\\
    &-\frac{(1\pm l)^2-r^2+z^2}{(1\pm l-r)^2+z^2}\frac{\sqrt{(1\pm l)m_\pm}}{2r^{3/2}}\Big(E(m_\pm) - (1-m_\pm)K(m_\pm)\Big)\Big],
  \end{align}
\begin{equation}\label{eq5.8}
    F_{\text{itr},z}(r,z;\pm l) =- \frac{GM}{2\pi R^2}\frac{z}{(1\pm l-r)^2+z^2}\sqrt{\frac{(1\pm l)m_\pm}{r}}E(m_\pm)\,,
\end{equation}
and
\begin{equation}\label{eq5.9}
S'_r=\frac{\partial S}{\partial r} =  N(r,z) \Big[\big(z^2 - (r-1)^2\big)
\frac{E(m_0)}{K(m_0)} -\frac{\big(z^4 - (r^2-1)^2\big)(1-m_0)}{4 r } 
 V(m_0)\Big]\, ,
\end{equation}
\begin{equation}\label{eq5.9}
 S'_z=\frac{\partial S}{\partial z} =  z\,N(r,z) \Big[2  (1-r)
\frac{E(m_0)}{K(m_0)} - \frac{1}{2} (r^2 - 1 + z^2)(1-m_0) 
 V(m_0)\Big]\, ,
\end{equation}
\begin{equation}
\displaystyle
   V(m_0) = 1 -2 \frac{E(m_0)}{K(m_0)}+\frac{1}{1-m_0}\frac{E(m_0)^2}{K(m_0)^2}\,, \qquad N(r,z)= \frac{1}{8\big((r-1)^2+z^2\big)^2}.
\end{equation}
The moduli of the elliptical integrals $m_0$ and $m_\pm$ are determined by the (\ref{eq3.4}) and (\ref{eq3.7a}). 

The (\ref{eq5.5}) and (\ref{eq5.6}) are the approximate expressions for the gravitational force components of a homogeneous torus with an elliptical cross-section. They are valid for both oblate and prolate cross-sections within the same limits of parameters given for the potential (see Section~\ref{section3a}). In the case of prolate tori, as in the case of the potential, the coordinates of \mcs \, are complex conjugate numbers. 
\begin{figure}[!h]
\centering
\includegraphics[width = 60mm]{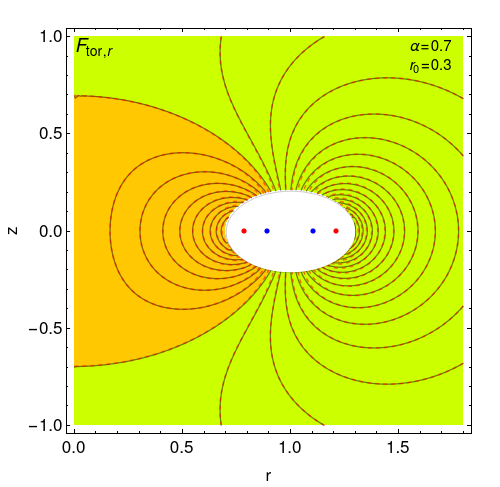}\qquad
\includegraphics[width = 60mm]{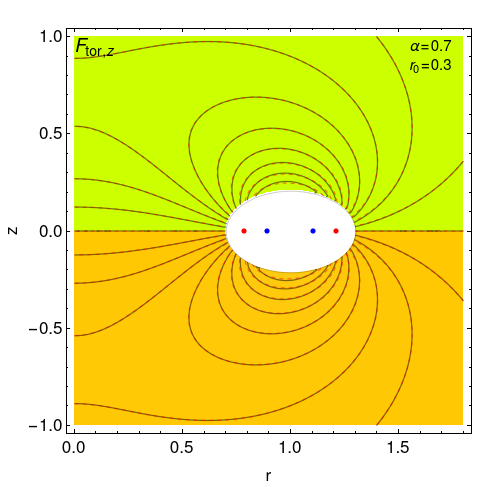}
\caption{The curves of the equal values for the radial ($F_{\text{tor},r}$) ({\it left panel}) and of the vertical ($F_{\text{tor},z}$) force component ({\it right panel}), calculated using the exact formulas (\ref{eq5.3})-(\ref{eq5.4})  (black) and the approximate expression (\ref{eq5.5}) - (\ref{eq5.5}) (red dashed), for an oblate cross-section ($r_0=0.3$, $\alpha = 0.7$). The colors divide the region with positive (orange) and negative (green) values of the force.}
\label{fig:force_oblate}
\end{figure} 
\begin{figure}[!h]
\centering
\includegraphics[width = 60mm]{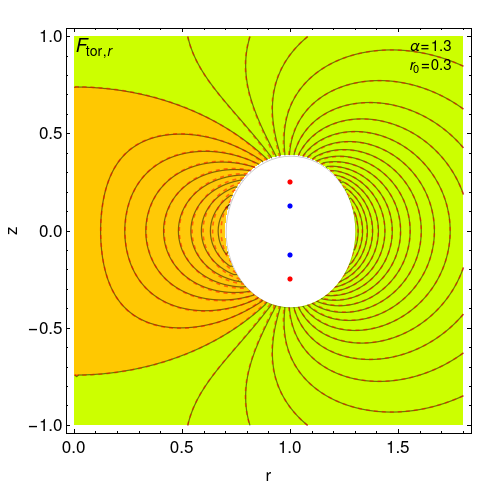}\qquad
\includegraphics[width = 60mm]{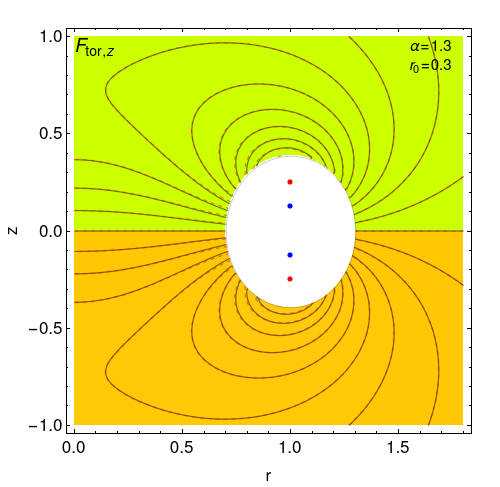}
\caption{The same as Fig.~\ref{fig:force_oblate} but for prolate cross-section ($r_0=0.3$, $\alpha = 1.3$).}
\label{fig:force_prolate}
\end{figure} 
The examples of equal force curves given in Figs.~\ref{fig:force_oblate}, \ref{fig:force_prolate} show the quality of agreement between the approximate expression (red dashed curves) and the exact one (black solid curve). 
The regions on the meridian plane $(x,z)$ where the radial component takes opposite signs (left panels in Fig. \ref{fig:force_oblate} and Fig. \ref{fig:force_prolate}) are separated by a dividing curve, $F_{\text{tor},r} = 0$. The radial component is also obviously zero along the $z$-axis.
The presence of these "dividing curves" is the reason why we cannot estimate the errors in the same way as we did it for the potential. 
Therefore we computed the modulus $\vert {\vec{F}}\vert = \sqrt{F_r^2+F_z^2}$ for both the exact and the approximate expressions of the force, and evaluated the relative errors in the same way as for the potential in the same range of the parameters. The mean relative error within the range of coordinates and parameters already used above never exceeds 1.5\% for both oblate and prolate cases. 

\newpage
\section{Conclusions}
\label{conclusion}
We have studied the outer potential of a torus with an elliptical (oblate/prolate) cross-section, both with homogeneous and inhomogeneous density distributions. 
Using a previous proven approach \citep{2011MNRAS.411..557B}, 
we have found  approximate expressions for the outer potential by representing the torus through a sum of the potentials of two \mcs. Our approach has yielded the following new properties.

\begin{itemize}
\item \label{item1} The outer potential of a homogeneous torus with an elliptical cross-section { is well represented}, with less than 1\% error, by the S$_{\alpha}$-approximation, i.e., by a combination of the potentials of two \mcs,  each with half the mass of the torus. These circles are located at half the focal length distances from the center of the cross-section along its major axis, irrespective of the size and flattening of the cross-section. 
\item At a first approximation, the ratio of the outer potentials of two confocal homogeneous tori with different masses coincides with the ratio of their masses.
\item The two \mc \, approach provides a simpler expression for the outer potential of an inhomogeneous torus with elliptical density contours in its cross-section than the integral expression. 
\item For a confocal density distribution, the outer potential of a torus with an elliptical cross-section is only weakly dependent on the density distribution law. Therefore, within the approximation limits, it is the same as that of a homogeneous torus and can be consequently represented by two \mc \, S$_{\alpha}$-approximation (see the first item of this list). 
\item We have also derived the components of the force for a homogeneous torus with an elliptical cross-section for both the exact and the approximate expressions of the potential. From their comparison we find that the modulus of the "approximate" force agrees with that of the "exact" force within 1.5\% within the range of coordinates where we also compared the potential.
\end{itemize}

\noindent
It is useful to note that our approach produced a unique solution for both  
oblate and prolate cross-sections.
It may simplify the investigations of the dynamics in the gravitational field of a torus, especially for analytical studies: the interpretation of N-body simulations for the central region of AGNs, the dynamics in the ring galaxies, and in the ring structures of Solar System bodies.

\bmhead{Acknowledgements}
\kolr{ The authors are grateful to the unknown referees for a critical reading of the manuscript and many useful suggestions.}
The work of EB and SS was supported under the special program of the NRF of Ukraine "Leading and Young Scientists Research Support" -- "Astrophysical Relativistic Galactic Objects (ARGO): life cycle of active nucleus", No.~2020.02/0346.

\bmhead{Data availability}

Simulation data and codes used in this paper can be made available upon request by emailing the corresponding author.

\begin{appendices}

\section{Comparison between the S-approximation and the series in toroidal coordinates}\label{secA1}
Here we compare two expansions into series of the outer gravitational potential of a homogeneous torus with circular cross-section.
One is the Maclaurin series up to the second-order terms in the vicinity of the central \mc, given by (\ref{eq3.1}), which we have called the S-approximation \cite{2011MNRAS.411..557B}.
The second one is the expansion of the potential in toroidal coordinates obtained by \cite{1973AnPhy..77..279W, Majic2020}, also limited to the second term: 
\begin{equation}\label{app_1.1}
\varphi_\text{tor} (r, z)=\frac{2}{3\pi^2} \frac{G M}{ R}  \frac{v^2}{r_0^2}  \Delta(r,z;\beta_0) \sum_{n=0}^\kolb{\infty} \epsilon_n C_n (\beta_0) P_{n-\frac{1}{2}}(\beta ) \cos (n\eta),    
\end{equation}
\kolr{ with}
\begin{equation}\label{app_1.2}
    C_n (\beta_0)= \Big(n + \frac{1}{2}\Big) Q_{n + \frac{1}{2}} (\beta_0)\, Q_{n - \frac{1}{2}}^2(\beta_0) - \Big(n - \frac{3}{2}\Big) Q_{n - \frac{1}{2}} (\beta_0) Q_{n + \frac{1}{2}}^2(\beta_0),
\end{equation}
\begin{equation}
  \gamma= \frac{1}{2}\ln{\frac{(r+v)^2 +z^2}{(r-v)^2 +z^2}}, \qquad
  \eta = \arccos\frac{r^2 + z^2 - v^2}{\sqrt{(r^2 + z^2 - v^2)^2 + 4v^2 z^2}}
\end{equation}
and
\begin{equation}\label{app_1.3}
    v = \sqrt{1-r_0^2}, \qquad \beta = \cosh\gamma,\qquad \beta_0 = 1/r_0=R/R_0, \qquad \Delta=\sqrt{2(\beta - \cos{\eta})}\,.
\end{equation}
\kolr{ The Neumann} factor is
\begin{equation}\label{app_1.4}
\epsilon_n=
 \begin{cases}
  1, n=0
   \\
  2, n\neq 0.
 \end{cases}  
\end{equation}
\begin{figure}[h]
\centering
\includegraphics[width = 60mm]{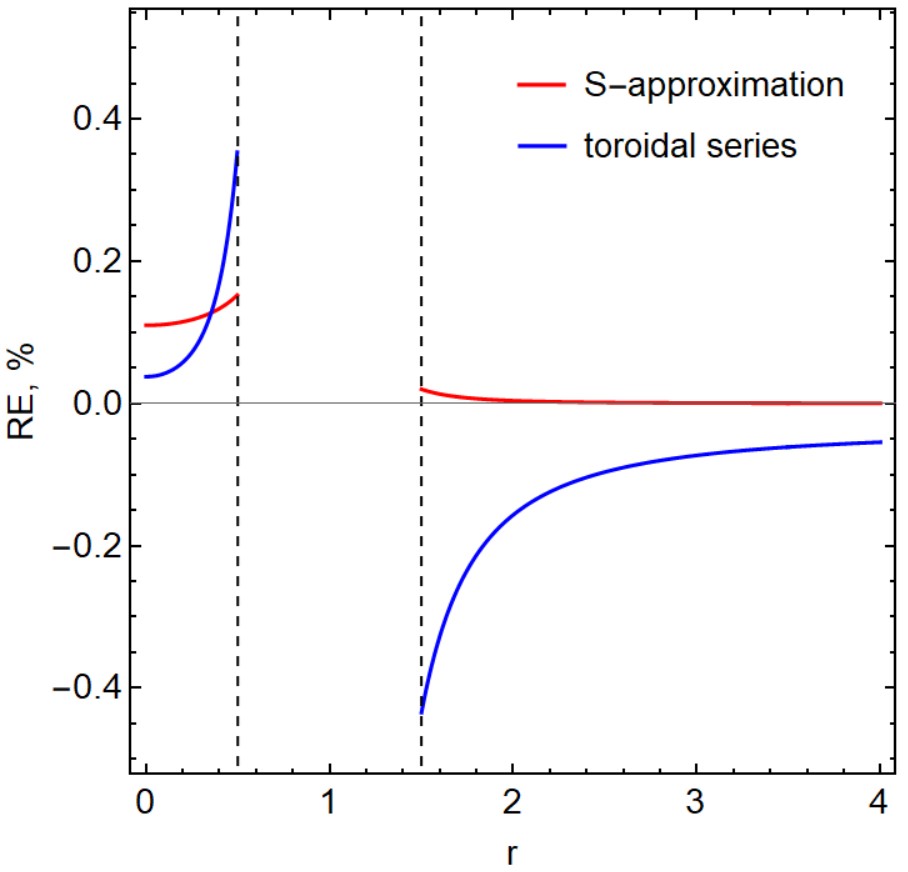}\quad
\includegraphics[width = 60mm]{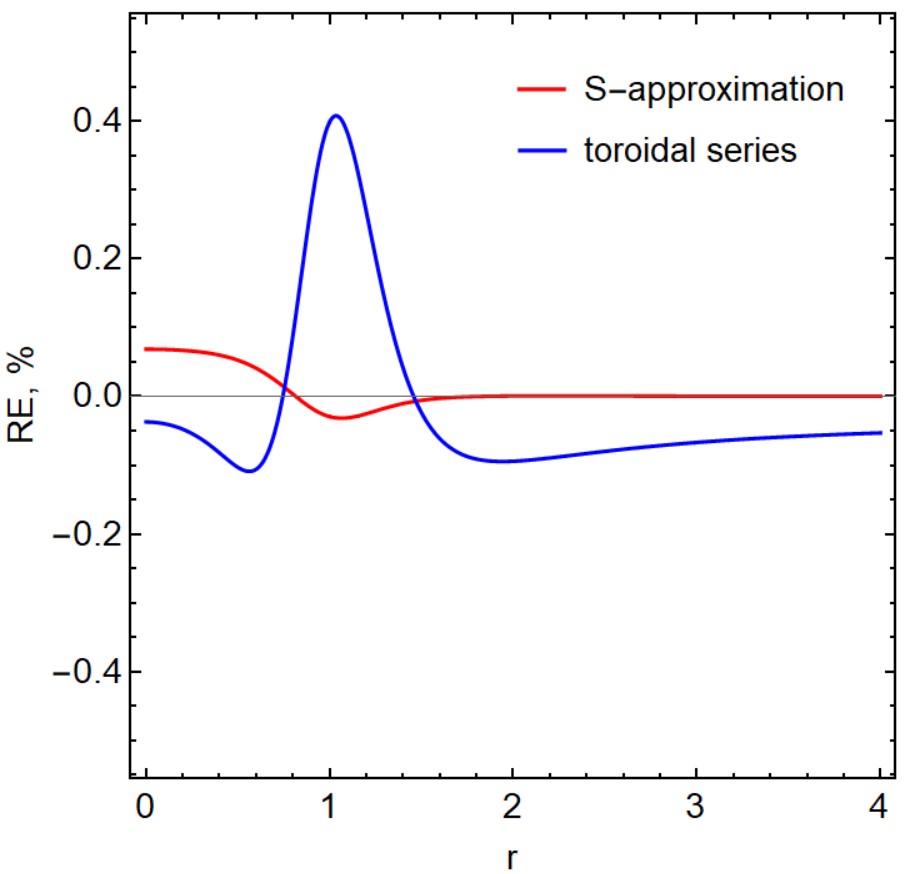}
\caption{\kolr{ Radial trends of the} relative errors for \kolr{ the} S-approximation (red) and for the \kolr{ expansion into series of the potential} in toroidal coordinates (blue) \kolr{ for the case of a  homogeneous circular torus ($r_0=0.5$)}, \kolr{ at} $z=0$ ({\it left}) and $z=0.5$ ({\it right}). { The dashed lines here and in the following figures mark the boundaries of the torus volume.}}
\label{fig:s-wong-comp}
\end{figure}
All coordinates are dimensionless and normalized to the major radius $R$ of the torus. 
Relative errors $\text{RE}=100\kolr{\times} (\varphi_\text{app} -\varphi_\text{tor})/\varphi_\text{tor}$ are presented in Fig.~\ref{fig:s-wong-comp}, where $\varphi_\text{tor}$ is the exact integral expression (\ref{eq2.2}) and $\varphi_\text{app}$ is the approximated expression for each of the two cases: S-approximation (\ref{eq3.1}) and toroidal series (\ref{app_1.1}). The comparison shows that in general the S-approximation has smaller relative errors
than the series in toroidal coordinates. We have chosen the case of a thick torus ($r_0=0.5$) as this is a more critical case than a thin torus.
\begin{figure}[h]
\centering
\includegraphics[width = 60mm]{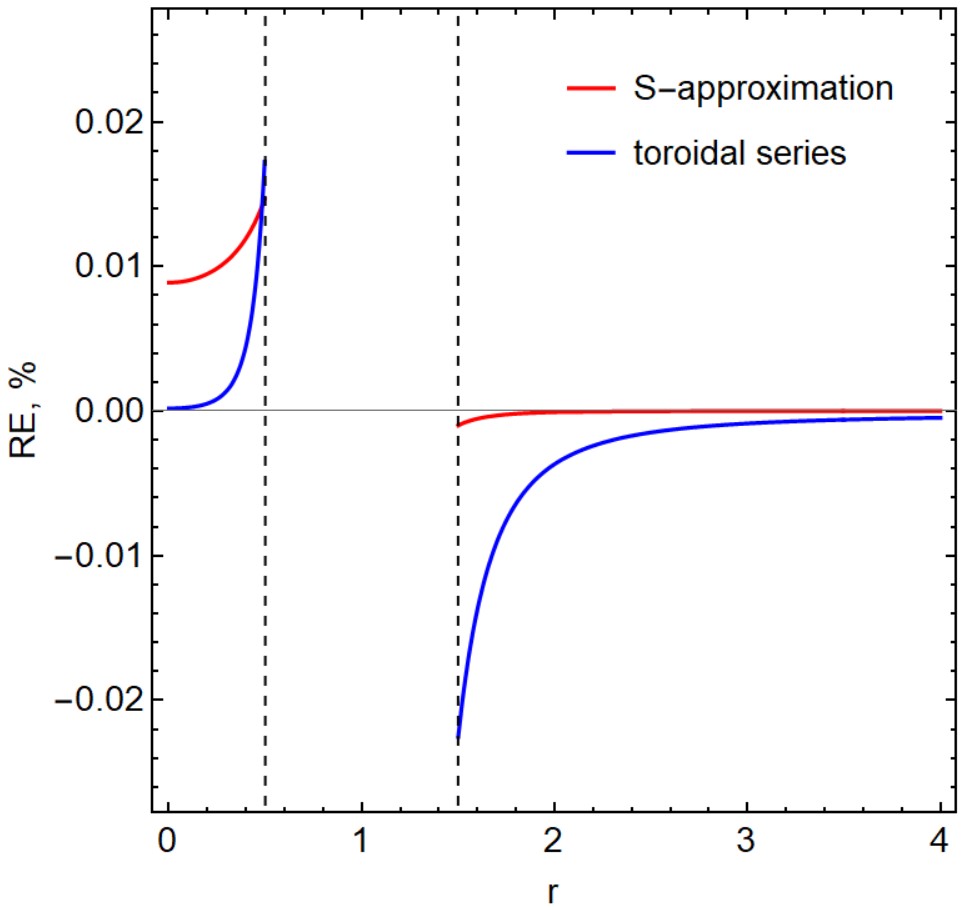}\quad
\includegraphics[width = 60mm]{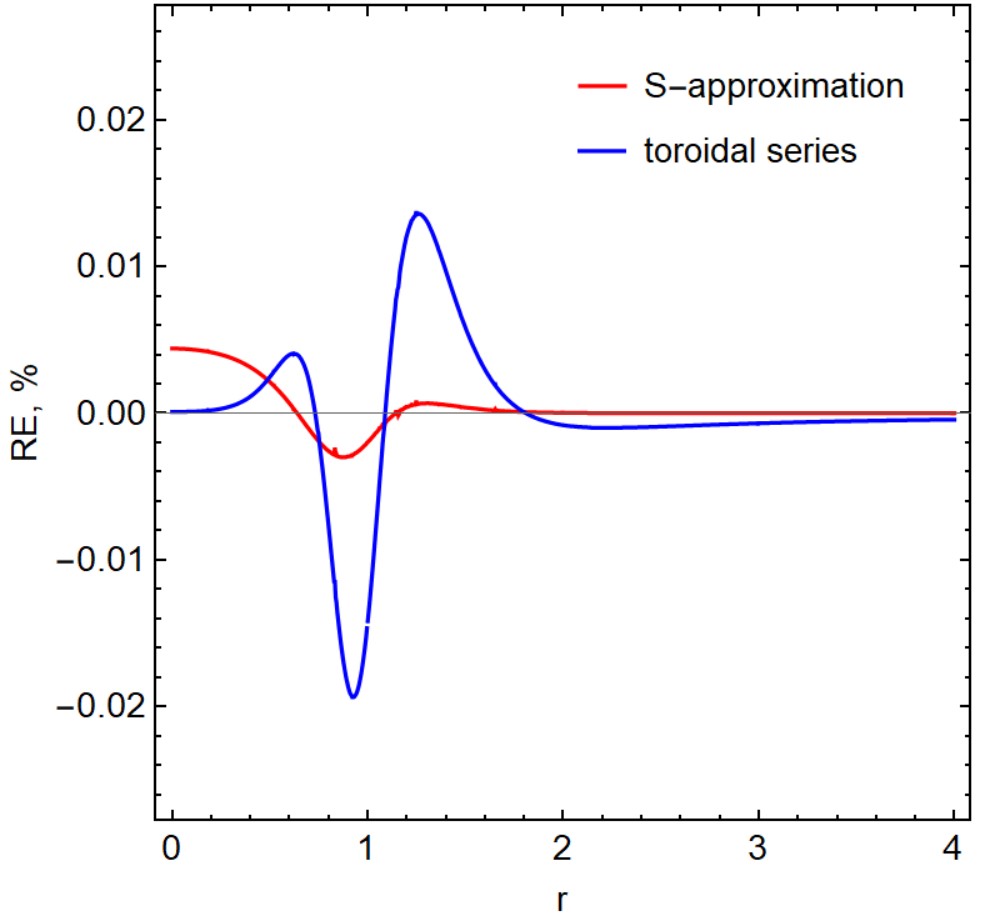}
\caption{\kolr{ Same as in Fig. \ref{fig:s-wong-comp} for an expansion of both series up to the 4-th order.}}
\label{fig:s-wong-comp-4th}
\end{figure}

Figure~\ref{fig:s-wong-comp-4th} shows the REs for the Maclaurin and toroidal series up to 4-th term. 
Note that for $r_0=0.5$, $z=0$, the 4-th term in the Maclaurin series is less than the second summand in (\ref{eq3.1}) by a factor of $\sim 10^{-2}$ in the torus hole and $\sim 10^{-3}$ outside. It is an order of magnitude less for the case of thin torus ($r_0=0.1)$. The series in toroidal coordinates gives the same order of differences in consecutive terms.

\section{Maclaurin series expansion up to the second term for the elliptical cross-section}\label{secA2}
\setcounter{figure}{0}  
\setcounter{equation}{0}
We consider here the Maclaurin series expansion (\ref{eq2.2}) of the integral expression for the torus potential with an elliptical cross-section. This is the same procedure that we used to obtain the outer potential of the torus with the circular cross-section \citep{2011MNRAS.411..557B}.
Since the function (\ref{eq2.3}) has no singularity for all values of $x', z'$, we can expand it in Maclaurin series in a neighbourhood of the central \mc \, ($x'=z'=0$). Restricting the expansion to the second order terms, we have
\begin{multline}\label{ap:2.1}
  \potmc(r,z;x',z') \approx
 \varphi_0(r,z) + 
 \left.\frac{\partial\potmc}{\partial x'}\right|_{\substack{x'=0 \\z'=0}} x'+ \left.\frac{\partial\potmc}{\partial z'}\right|_{\substack{x'=0 \\z'=0}} z'+ \\*
       \kolr{+} \frac{1}{2} \Big(\left.\frac{\partial^2\potmc}{\partial x'^2}\right|_{\substack{x'=0 \\z'=0}} x'^2+ \left.\frac{\partial^2\potmc}{\partial z'^2}\right|_{\substack{x'=0 \\z'=0}}z'^2+
       2\left.\frac{\partial^2\potmc}{\partial x'\partial z'}\right|_{\substack{x'=0 \\z'=0}}x'z'\Big)\,,
\end{multline}  
where and $\varphi_0(r,z)$ is the potential of the central \mc \, (\ref{eq3.2}). To obtain the potential of the torus, we substitute (\ref{ap:2.1}) in expression (\ref{eq2.2}). After integration, odd-power terms of coordinates disappear since the integrals have symmetric limits. Only even-power terms differ from zero:
\begin{multline}\label{ap:2.1b}
  \varphi_\text{tor}(r,z) \approx
 \varphi_0(r,z) \int\limits_{-r_0}^{r_0} 
        \int\limits_{-z_e}^{z_e}dz'dx'+ 
       \frac{1}{2} \Big(\left.
            \frac{\partial^2\potmc}{\partial x'^2}
      \right|_{\substack{x'=0 \\
                        z'=0}}  \times\int\limits_{-r_0}^{r_0}\int\limits_{-z_e}^{z_e}x'^2dz'dx'+ \\
       +\left.
                     \frac{\partial^2\potmc}{\partial  z'^2}
                    \right|_{\substack{x'=0 \\
                                    z'=0}}\times\int\limits_{-r_0}^{r_0}\int\limits_{-z_e}^{z_e}z'^2dz'dx'\Big)\,.
\end{multline} 
\begin{figure}[h!]
\centering
\includegraphics[width = 60mm]{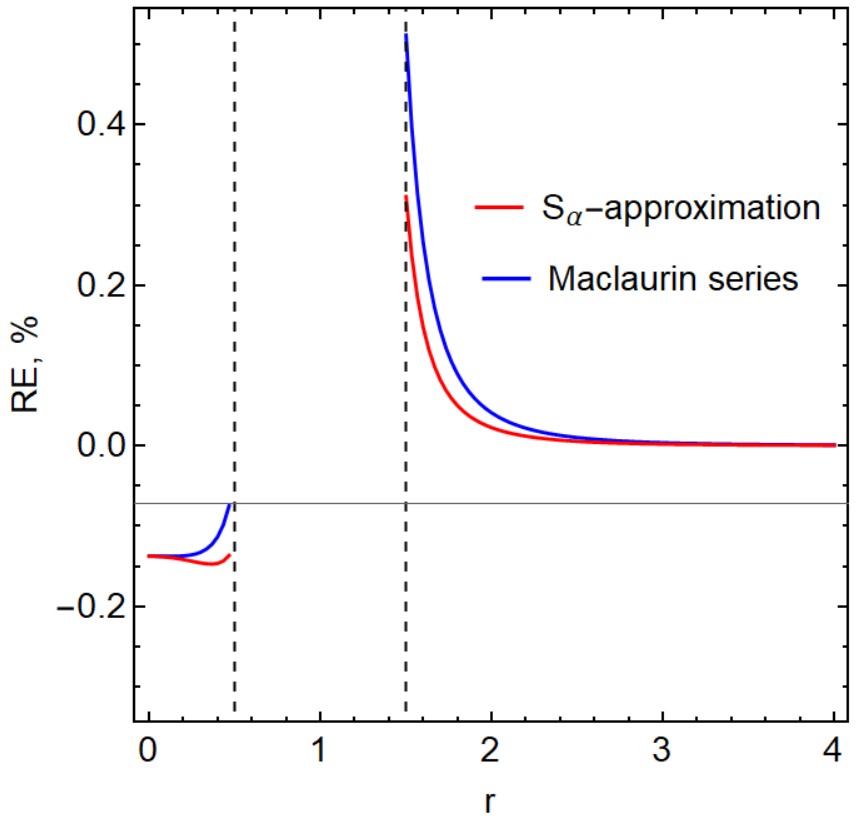}\quad
\includegraphics[width = 60mm]{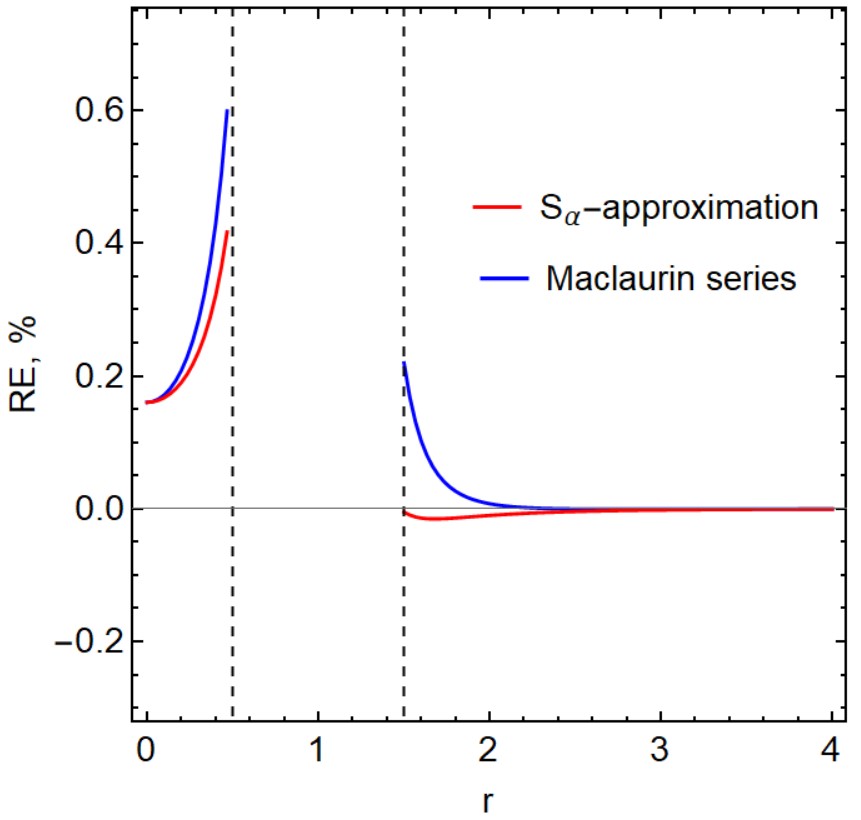}
\includegraphics[width = 60mm]{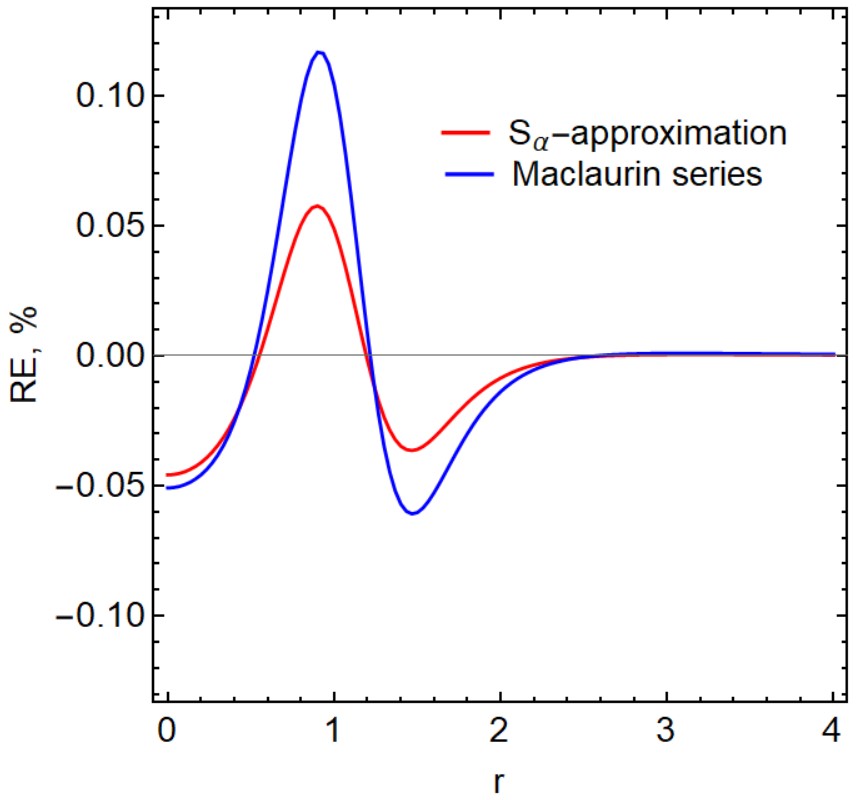}\quad
\includegraphics[width = 60mm]{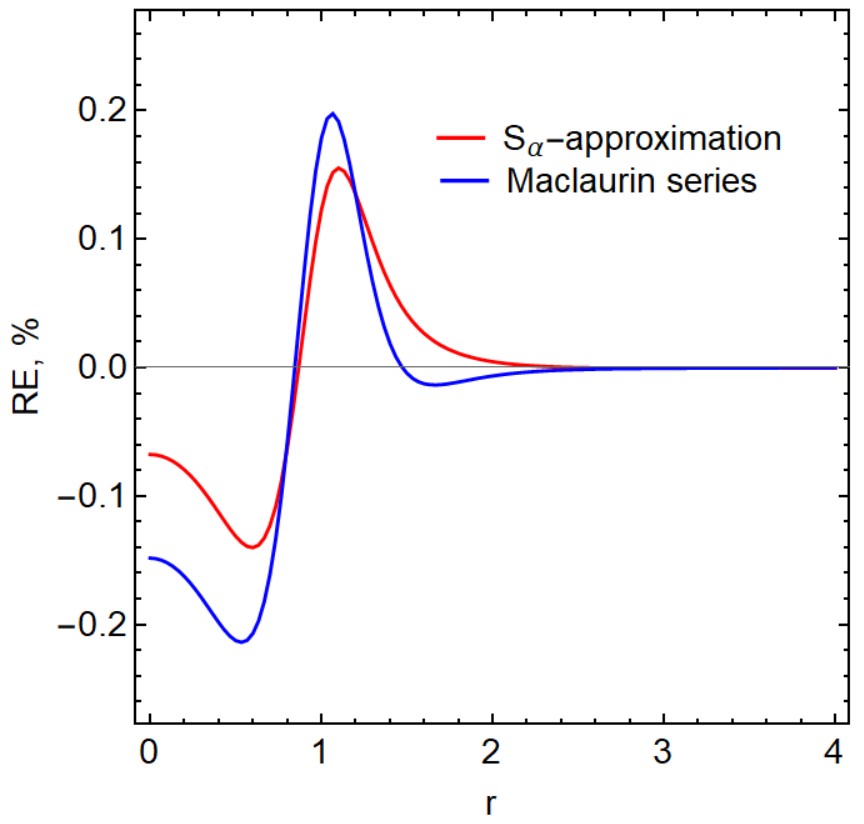}
\caption{Relative errors for the two \mc \, approximation (\ref{eq3.5}) (red) and for the Maclaurin series expansion (\ref{ap:2.4}) (blue)  for \kolr{ a} torus thickness $r_0=0.5$: $z=0$ ({\it top \kolr{ panels}}) and $z=0.7$ ({\it bottom \kolr{ panels}}), $\alpha=0.7$ ({\it left \kolr{ panels}}; oblate case) and $\alpha=1.3$ ({\it right \kolr{ panels}}; prolate case). }
\label{fig:Maclaurin-twor-comp}
\end{figure}
When we take the integrals and $\varphi_0(r,z)$ out of the brackets, we have:
\begin{equation}\label{ap:2.4}
\varphi_\text{tor}(r,z) \approx \varphi_0(r,z) 
 \left[1 + \frac{r_0^2}{8}(A+ \alpha^2 B)\right]~,
\end{equation}  
where 
\begin{equation}\label{ap:2.5}
A(r, z) = \left. \frac{1}{\varphi_\text{0}}
            \frac{\partial^2\potmc}{\partial x'^2}
      \right|_{\substack{x'=0 \\
                        z'=0}},
\qquad \qquad B(r, z) = \left.
                     \frac{1}{\varphi_\text{0}}\frac{\partial^2\potmc}{\partial  z'^2}
                    \right|_{\substack{x'=0 \\
                                    z'=0}}\,.
\end{equation}
After simplifications, the expressions for $A$ and $B$ have the following form: 
\begin{multline}\label{ap:2.6}
A=\frac{C}{2}\bigg[\Big(z^6 + d_-^2d_+ +  z^4(3r^2-5)
+ z^2(3r^4 - 6r^2 - 5)\Big)\frac{E(m_0)}{K(m_0)}- z^6 \kolr{-} \\ -z^4(3r^2-2r+1) - z^2(3r^4 - 4r^3 + 1) - p_-^4p_+^2\bigg],
 \end{multline}

\begin{equation}\label{ap:2.7}
B = C\bigg[\Big(3z^4- d_-^2 + 2 z^2d_+\Big)\frac{E(m_0)}{K(m_0)} -z^2\Big(z^2 + p_-^2\Big)\bigg],
\end{equation}
with
\begin{equation}\label{ap:2.8}
C = \frac{1}{(z^2 + p_-^2)^2(z^2 + p_+^2)},
\end{equation}
\\
and $p_\pm = r \pm 1$, $d_\pm = r^2 \pm 1$.
In the limit case $\alpha = 1$,  i.e. for a torus with circular cross-section, the expression (\ref{ap:2.4}) turns into  the
S-approximation (\ref{eq3.1}). Figure \ref{fig:Maclaurin-twor-comp} shows the comparison of this case with the  \mbox{S$_\alpha$-approximation}. 

%----------------------------------------------------------------------------
\section{Testing the positions of the \mcs\ in the S$_{\alpha}$-approximation}\label{secA2add}
\setcounter{figure}{0}  
\setcounter{equation}{0}
Here we investigate numerically whether $l=f/2$ is the best choice for the distances of the two \mcs \, from the center of the elliptical cross-section in the S$_\alpha$-approximation for the outer potential of a homogeneous torus. 

Let then assume that the two distances \kolr{ may be} different: $l_1=k_1 f$, $l_2=k_2 f$, with the focal length $f= r_0 \sqrt{1 - \alpha^2}$. The gravitational potential writes as:
\begin{equation}\label{eq_appA2add:1}
\varphi^\text{app}_\text{tor}  (r,z)= 
             \Big(\potmc(r,z;-l_1)+\potmc(r,z;+l_2)\Big)
            \Big(1 - \alpha^2r_0^2 S(r,z)\Big),
\end{equation}
The potential of the each \mc \, ($\potmc$) is provided by the expression (\ref{eq3.6}) with $l$ replaced by $l_1$ and $l_2$, respectively.
\begin{figure}[h]
\centering
\includegraphics[width = 60mm]{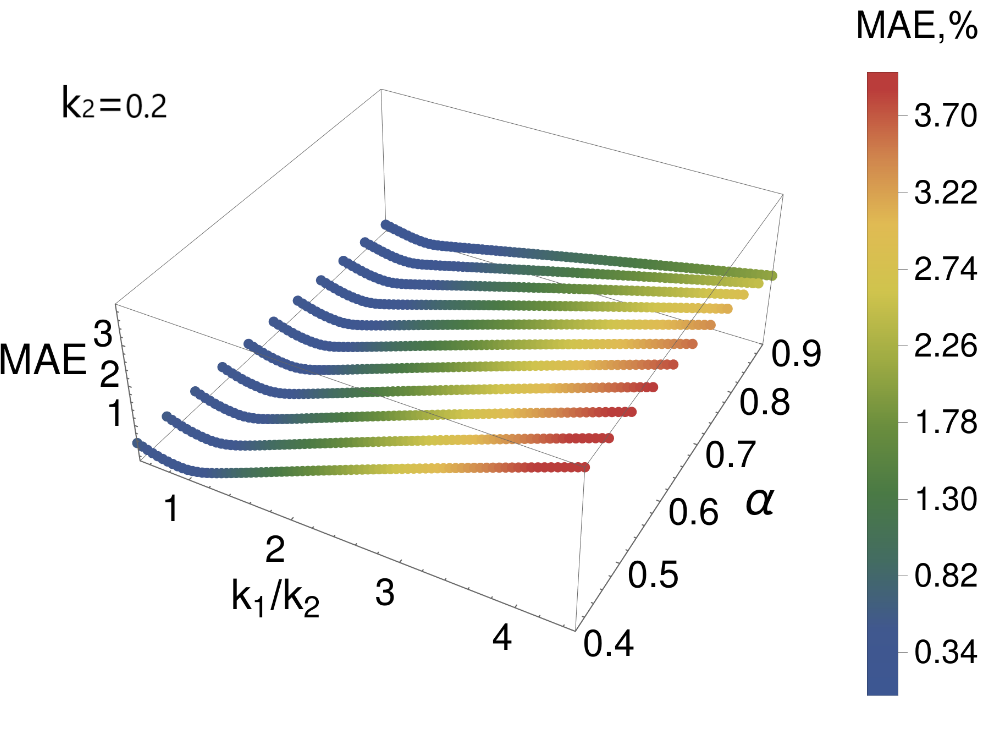}\qquad
\includegraphics[width = 60mm]{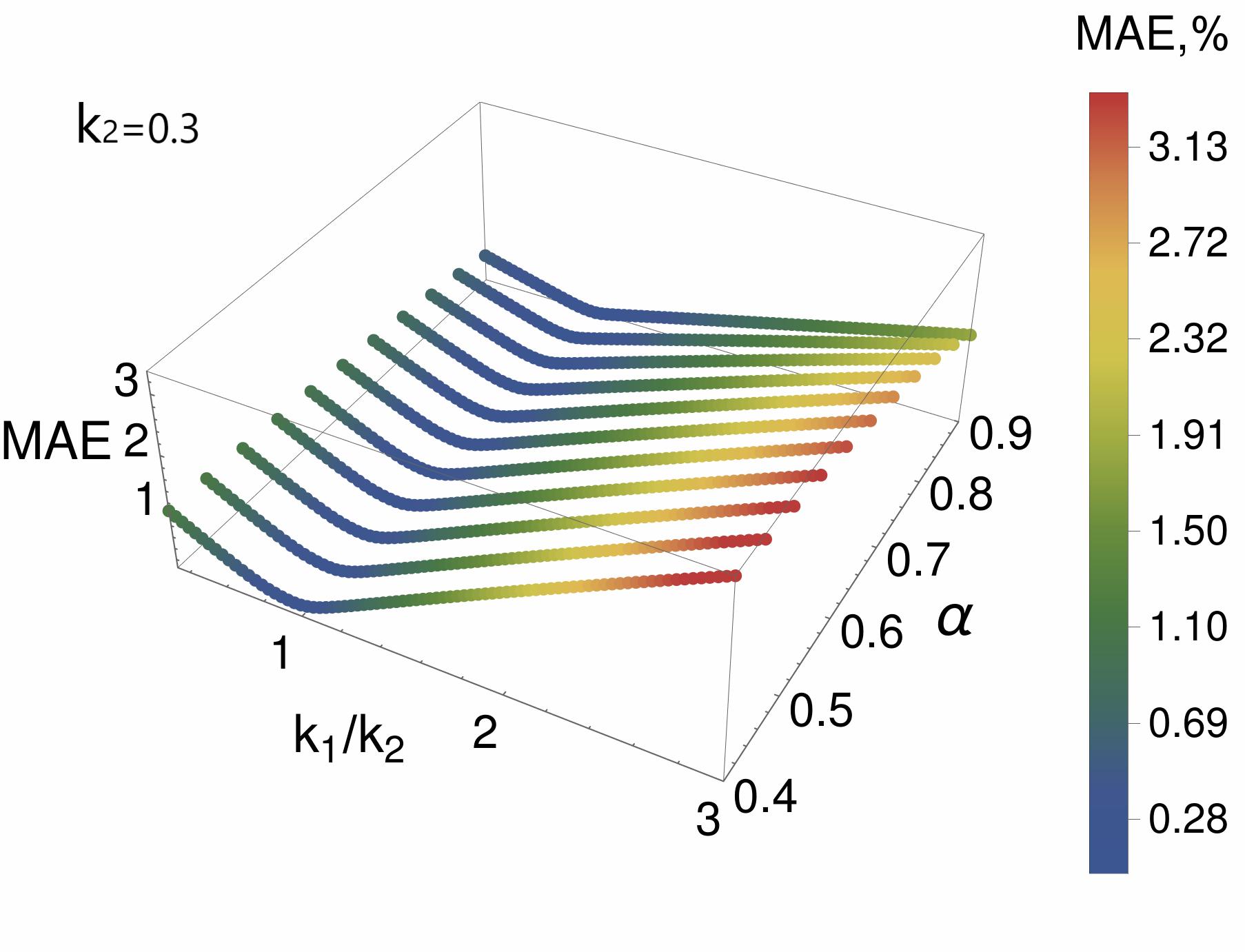}
\includegraphics[width = 60mm]{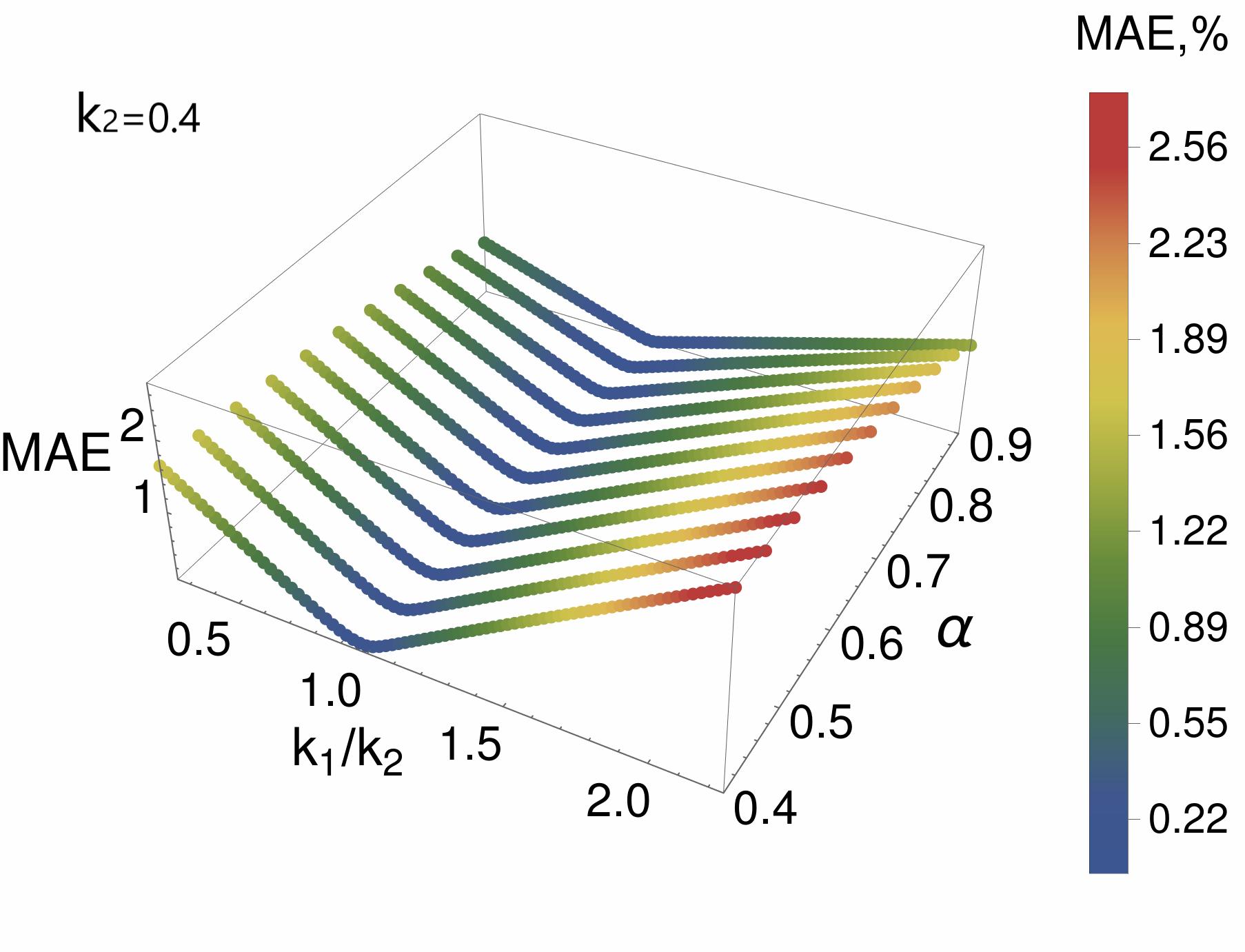}\qquad
\includegraphics[width = 60mm]{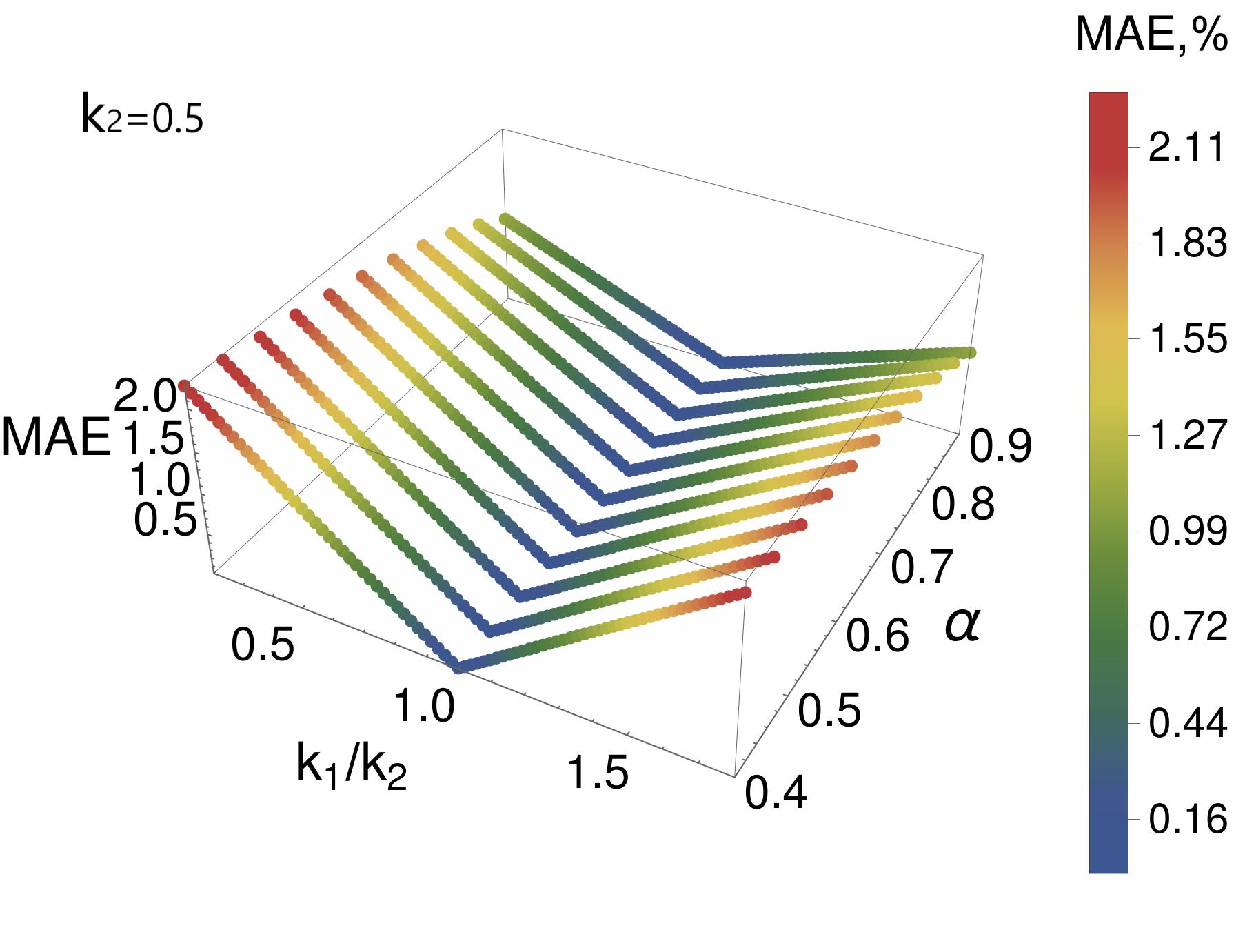}
\includegraphics[width = 60mm]{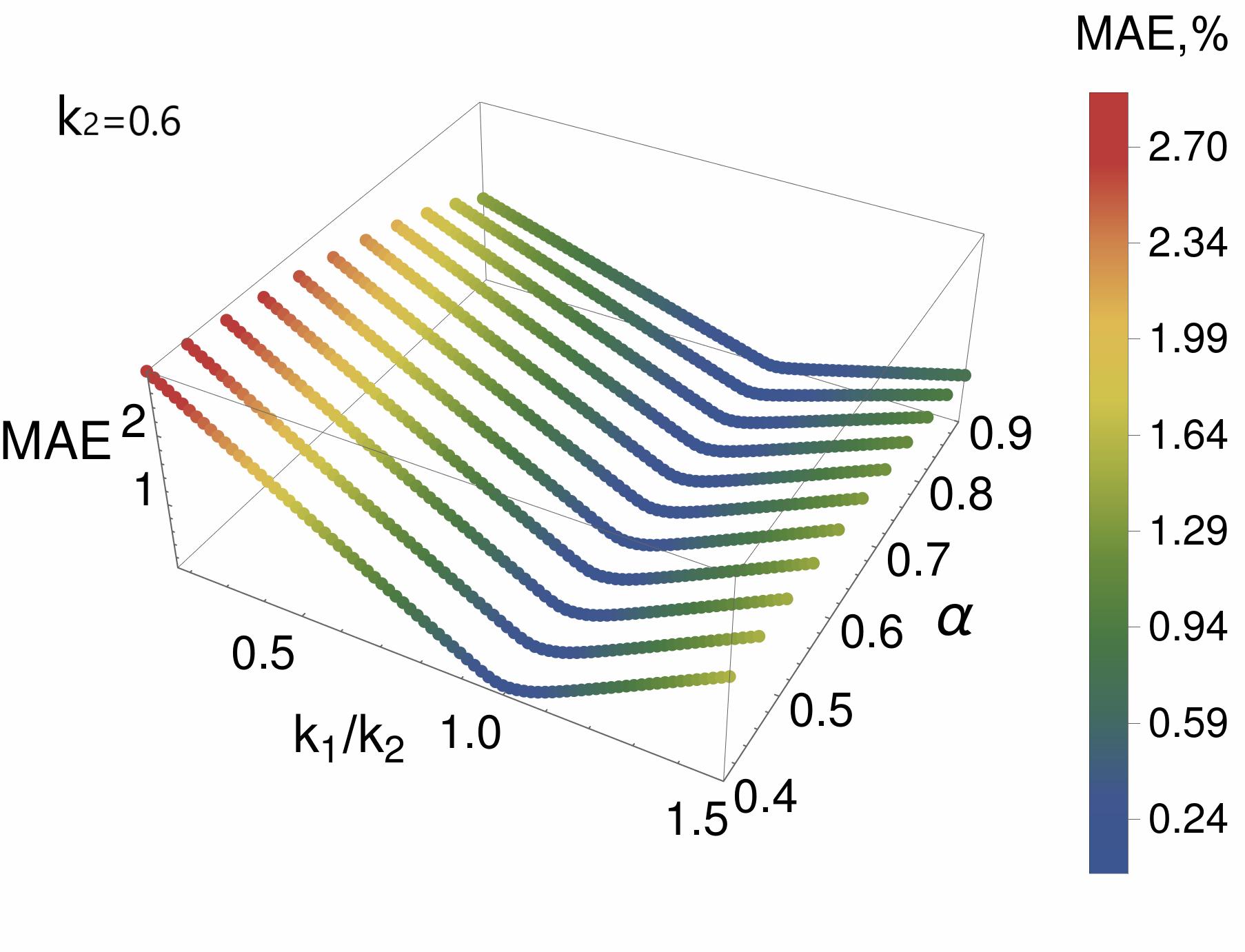}\qquad
\includegraphics[width = 60mm]{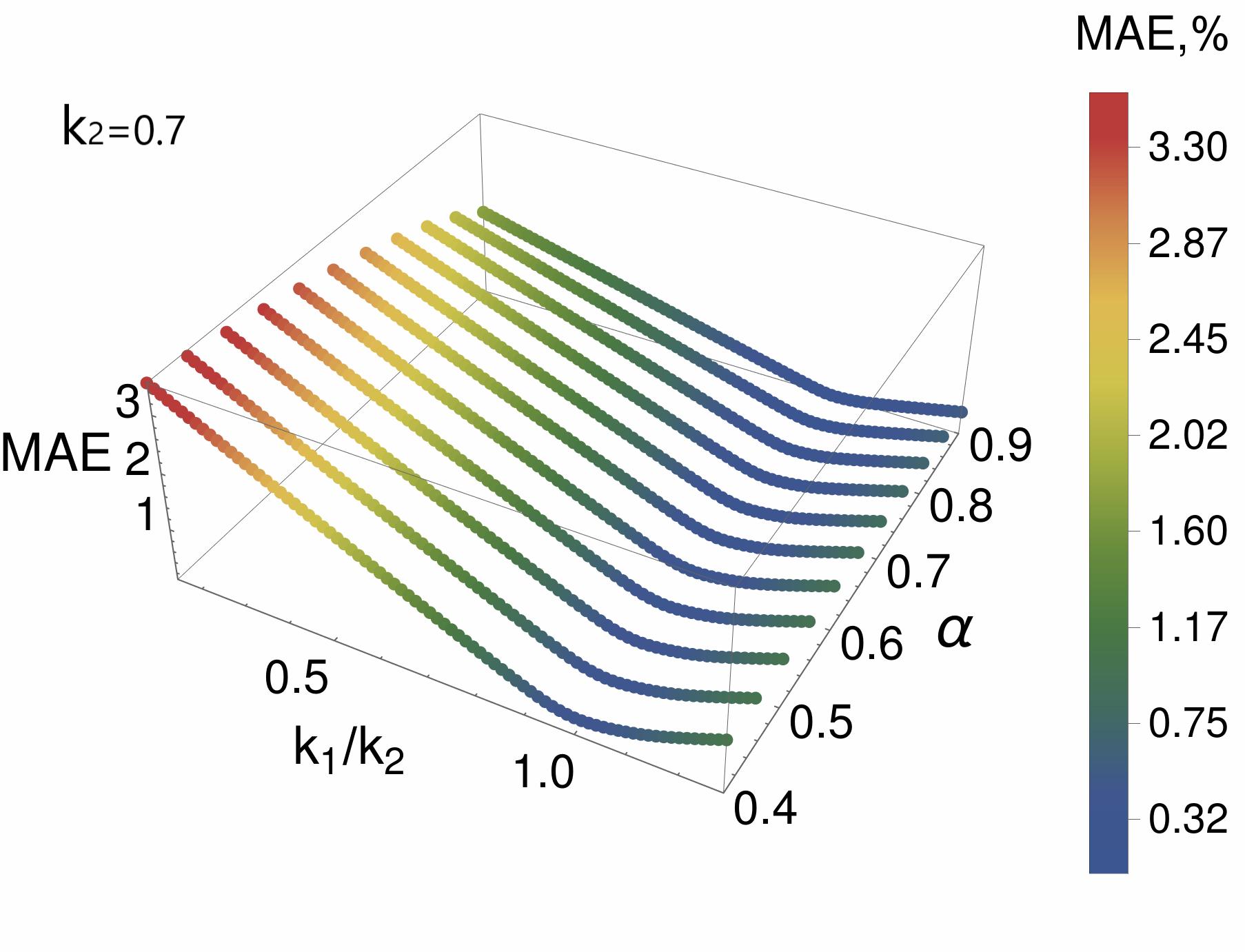}
\caption{\kolr{ Mean absolute deviations (MAE) of the \kolb{S$_\alpha$-approximation} from the exact formula for the potential of a homogeneous torus with an elliptical cross-section for the different values of the distance ratios $k_1/k_2$, keeping $k_2$ fixed. The parameter $k_1$ spans the interval $[0.1,0.9]$. The smallest errors occur for $k_1=k_2=0.5$.}}
\label{fig:k1_k2}
\end{figure}

Figure~\ref{fig:k1_k2} shows examples of the mean average deviations (same as (\ref{3.7b})) of the $S_\alpha$-approximation from the exact formula for different pairs of values of the distances $l_1$ and $l_2$. To explore the parameter space, we have fixed one of the 
distance coefficients, $k_2$, and computed the MAE by varying the cross-section ellipticity $\alpha$ and the ratio $k_1/k_2$. In all cases the smaller values of MAE occur at $k_1/k_2=1$, i.e., when the two distances are equal. The errors are lowest when the distance $l_1=l_2=l=f/2$.

\newpage
\section{The complex coordinates of \mcs\ for prolate cross-section}\label{secA3}
\setcounter{figure}{0}  
\setcounter{equation}{0}
Formula (\ref{eq3.5}) also works for prolate cross-sections ($\alpha > 1$), but we have to use the imaginary conjugate numbers for the coordinates of the \mcs. To understand it, let us recall the Cartesian representation of a complex number, with the real part on the abscissa axis and imaginary coefficient on the ordinate. So, when the elliptical cross-section of the torus becomes prolate, the foci lie formally on the imaginary axis. As the result, the coordinates of the  \mcs \, also became imaginary and conjugate, but the resulting values of the potential remain real.
Let us proof it. 

Indeed, we can represent the coordinates of the \mcs \, as: 
\begin{equation}\label{ap:3.1}%\label{eq4.6}
l= \pm i y.
\end{equation}
Substituting (\ref{ap:3.1}) in the expression (\ref{eq3.6}), we have
\begin{equation}\label{eqDD1}
\potmc(r,z;\pm iy) = \frac{G M}{2\pi R}\frac{2(1 \pm iy)}{\sqrt{(1\pm iy+r)^2 + z^2}}  \, K(m_\pm),
\end{equation}
where the parameters of the elliptical integrals (\ref{eq3.7a}) are:
\begin{equation}\label{eqDD2}
\displaystyle m_\pm = \frac{4 r \, (1 \pm iy)}{(1 \pm iy+ r)^2 + z^2}~.
\end{equation}
After simplifications, we can rewrite the expression (\ref{eqDD2}) as follows:
\begin{equation}\label{ap:3.2}%\label{eq4.7a}
  m_\pm = \beta \pm i \gamma,  
\end{equation}
where
\begin{equation}\label{ap:3.3}%\label{eq4.8}
\beta = \frac{4r \, (\xi + 2(1+r)z^2)}{\xi^2 + 4(1+r)^2 z^2},
\qquad
\gamma = \frac{4r \, y\, (\xi - 2(1+r))}{\xi^2 + 4(1+r)^2 y^2},
\qquad
\xi = (1+r)^2 + z^2 - y^2.
\end{equation}
It is seen that the parameters are conjugate: $m_- = \overline{m_+}$.
The elliptical integrals \kolg{ for} conjugate parameters also give conjugate numbers that we can write in the form: 
\begin{equation}\label{ap:3.6} 
  K(m_\pm) = \lambda \pm i \sigma. 
\end{equation}
Let's consider the multiplier in (\ref{eqDD1}) \kolr{ separately}:
\begin{equation}\label{ap:3.7} 
   L_{\pm}= \frac{2(1 \pm iy)}{\sqrt{(1\pm iy+r)^2 + z^2}}\,.
\end{equation}
After simplifications and taking into account the notations (\ref{ap:3.3}), we have
\begin{equation}\label{ap:3.7b} 
   L_{\pm}= 
  \sqrt{\frac{1}{r}\Big[{(\beta -\gamma z) \pm 
   i (\beta z + \gamma)\Big]}}\,.
\end{equation}
To proof that $L_\pm$ \kolr{ corresponds to} conjugate numbers, we represent it in the form: 
\begin{equation}\label{ap:3.8} 
    L_{\pm} = \mu \pm i \nu\,.
\end{equation}
Squaring (\ref{ap:3.7b}) and (\ref{ap:3.8}), we obtain the equation  corresponding to the sign in the index. The comparison of the real and imaginary parts gives us the system of the equations: 
\begin{align}
  & 
  \frac{1}{r}\Big(\beta - \gamma z\Big) = \mu^2 - \nu^2, \label{ap:3.9a} \\ 
   & \frac{1}{r}\Big(\beta z + \gamma\Big) =  2\mu\nu . \label{ap:3.9b} 
\end{align}
By expressing $\mu$ from the equation (\ref{ap:3.9b}) and putting it in (\ref{ap:3.9a}), we obtain:
\begin{equation}\label{ap:3.10} 
\nu^4 + \frac{\beta - \gamma z}{r}\nu^2 - \frac{(\beta y + \gamma)^2}{4r^2}=0.
\end{equation}
We have only one real and positive root of this equation, that is:
\begin{equation}\label{ap:3.11} %\label{eq4.11ff}
\nu = \sqrt{\frac{1}{2r}
\Big[ \sqrt{\beta^2 + \gamma^2 + (\gamma z)^2 + (\beta z)^2}+\gamma z - \beta \Big]},
\end{equation}
and
\begin{equation}\label{ap:3.11} 
\mu = \frac{(\beta z+\gamma)}{2r\nu}.
\end{equation}
Note that the negative root of (\ref{ap:3.10})  also satisfies  the conjugation condition (\ref{ap:3.8}) by only changing the sign of $\nu$.
Replacing (\ref{ap:3.6}) and (\ref{ap:3.8}) in the sum of  the dimensionless potentials of the two \mcs \, (the second multiplier in (\ref{eq3.5})), we have:
 \begin{equation}\label{ap:3.12} 
 \potmc(+l) + \potmc(-l) =  2(\mu\lambda - \nu\sigma),
 \end{equation}
where $\mu$ and $\nu$ are functions of the coordinates.
So, the resulting quantity is a real one. Since the last multiplier in the expression (\ref{eq3.5}) is also real, the final value of the potential is real.          

\end{appendices}

\bibliography{bibliography}

\end{document}